\documentclass[12pt,a4paper]{article}
\usepackage{amsmath,amssymb,cite,slashed,tabularx}
\usepackage{subfigure}
\usepackage{graphicx,color}

\allowdisplaybreaks

\newcommand{\TeV}{\,\textrm{TeV}}
\newcommand{\GeV}{\,\textrm{GeV}}
\newcommand{\MeV}{\,\textrm{MeV}}

\numberwithin{equation}{section} 
\def\beq#1\eeq{\begin{align}#1\end{align}}

\newcommand{\CP}{${C\hspace{-0.2mm}P}\hspace{-0.4mm}~$}
\newcommand{\CPC}{${C\hspace{-0.2mm}P}$-conserving~}
\newcommand{\CPV}{${C\hspace{-0.2mm}P}$-violating~}
\newcommand{\epoe}{\ensuremath{\varepsilon'/\varepsilon_K}}
\newcommand{\epsk}{\ensuremath{\varepsilon_K}}

\usepackage{caption}
\definecolor{BlueViolet}{rgb}{0.2, 0.00, 0.7}
\definecolor{Blue}{rgb}{0.15, 0.00, 0.9}
\usepackage[colorlinks=true,linkcolor=Blue,citecolor=Blue,urlcolor=BlueViolet]{hyperref}

\begin{document}

\begin{titlepage}

\begin{center}

\hfill KEK--TH--2021\\
\hfill TTP17--050\\
\hfill December 2017

\vskip .35in

{\Large\bf
Gluino-mediated electroweak penguin \\ \vspace{2mm}
with flavor-violating trilinear couplings
}

\vskip .4in

{\large
  Motoi Endo$^{\rm (a,b)}$, 
  Toru Goto$^{\rm (a)}$, 
  Teppei Kitahara$^{\rm (c,d)}$,
  Satoshi Mishima$^{\rm (a)}$,\\ 
\vspace{.2cm}}
{\large 
  Daiki Ueda$^{\rm (b)}$,
 and
  Kei Yamamoto$^{\rm (e,f)}$
}

\vskip 0.25in

$^{\rm (a)}${\it
Theory Center, IPNS, KEK, Tsukuba, Ibaraki 305-0801, Japan}

\vskip 0.1in

$^{\rm (b)}${\it
The Graduate University of Advanced Studies (Sokendai),\\
Tsukuba, Ibaraki 305-0801, Japan}

\vskip 0.1in

$^{\rm (c)}${\it 
Institute for Theoretical Particle Physics (TTP), Karlsruhe Institute of Technology, Engesserstra{\ss}e 7, D-76128 Karlsruhe, Germany}
 
\vskip 0.1in
 
$^{\rm (d)}${\it
Institute for Nuclear Physics (IKP), Karlsruhe Institute of
Technology, Hermann-von-Helmholtz-Platz 1, D-76344
Eggenstein-Leopoldshafen, Germany} 

\vskip 0.1in
 
$^{\rm (e)}${\it
Department of Physics, Nagoya University, Nagoya, Aichi 464-8602, Japan}

\vskip 0.1in
 
$^{\rm (f)}${\it
Kobayashi-Maskawa Institute for the Origin of Particles and the Universe (KMI),\\
Nagoya University, Nagoya, Aichi 464-8602, Japan}

\end{center}

\vskip .3in

\begin{abstract}
In light of a discrepancy of the direct \CP violation in $K\to\pi\pi$ decays, \epoe, we investigate gluino contributions to the electroweak penguin, where flavor violations are induced by squark trilinear couplings.
Top-Yukawa contributions to $\Delta S = 2$ observables are taken into account, and vacuum stability conditions are evaluated in detail. 
It is found that this scenario can explain the discrepancy of \epoe\,for the squark mass smaller than $5.6\TeV$.
We also show that the gluino contributions can amplify $\mathcal{B}(K \to \pi \nu \overline{\nu})$, $\mathcal{B}(K_S \to \mu^+ \mu^-)_{\rm eff}$ and $\Delta A_{\rm CP}(b\to s\gamma)$. 
Such large effects could be measured in future experiments.
\end{abstract}
\end{titlepage}

\setcounter{page}{1}
\renewcommand{\thefootnote}{\#\arabic{footnote}}
\setcounter{footnote}{0}

\hrule
\tableofcontents
\vskip .2in
\hrule
\vskip .4in

\section{Introduction}

Flavor-changing neutral current (FCNC) processes are sensitive probes for new physics beyond the Standard Model (SM), since in the SM there are no FCNC processes at the tree level, and they are suppressed further by the Glashow-Iliopoulos-Maiani (GIM) mechanism. One of the FCNC observables, the \CPV ratio \epoe\,in neutral kaon decays into two pions, has attracted attentions recently because of a discrepancy between the experimental data and the theoretical predictions based on the first lattice calculation of the hadronic parameters $B_6^{(1/2)}$ and $B_8^{(3/2)}$ by the RBC-UKQCD collaboration~\cite{Bai:2015nea,Blum:2011ng, Blum:2012uk, Blum:2015ywa}.\footnote{
 In contrast, the chiral perturbation theory predicts  $B_6^{(1/2)} \approx 1.5$, which is a relatively larger value than the lattice result, and a consistent value with the measured \epoe\,is predicted~\cite{Pallante:2000hk,Pallante:2001he,Hambye:2003cy,Mullor}.}
The next-to-leading order (NLO) prediction for \epoe\,has been calculated in Ref.~\cite{Buras:2015yba}, and it has been confirmed by an improved calculation in Ref.~\cite{Kitahara:2016nld}. 
The latter result is given by 
\begin{align}
\left(\epoe\right)^{\mathrm{SM}}
=
(1.06 \pm 5.07) \times 10^{-4}, 
\end{align}
which deviates from the experimental data~\cite{Batley:2002gn,AlaviHarati:2002ye,Abouzaid:2010ny,Olive:2016xmw}  
\begin{align}
\textrm{Re}\left(\epoe\right)^{\mathrm{exp}}
= (16.6 \pm 2.3) \times 10^{-4} ,
\end{align}
at the $2.8\,\sigma$ level. The theoretical result which is much smaller than the data is supported by analyses in the large-$N_c$ dual QCD approach~\cite{Buras:2015xba,Buras:2016fys}. Note that improvements of the lattice calculation and independent confirmations of the result by other lattice collaborations are highly important to establish the presence of new physics in \epoe. 

In this paper, we study \epoe\,in the minimal supersymmetric standard model (MSSM) with introducing large off-diagonal entries in the trilinear couplings of the down-type squarks to the Higgs boson. 
The off-diagonal couplings generate gluino contributions to the flavor-changing $Z$ penguin which affects \epoe\,via the $I=2$ amplitude. 
Although such a scenario has been studied in Ref.~\cite{Tanimoto:2016yfy}, top-Yukawa contributions to $\Delta F = 2$ observables have not been taken into account.
In the scenario, \epsk\,receives those contributions from the $Z$ penguin through the renormalization group (RG) running from the new physics scale to the electroweak (EW) scale, and through the matching onto the low-energy FCNC operators at the EW scale~\cite{Endo:2016tnu,Bobeth:2017xry}. 
They can be comparable in size to ordinary gluino box contributions.
Moreover, since the LHC experiment is pushing up the lower bounds on the squark and gluino masses~\cite{Sirunyan:2017cwe,ATLAS}, the situation changes: larger trilinear couplings are required to explain the $\varepsilon'/\varepsilon_K$ discrepancy.

The large off-diagonal trilinear couplings also affect other FCNC observables. We consider constraints on the couplings as well as on other MSSM parameters from the branching ratios of $K_L\to \mu^+\mu^-$, $\bar{B}\to X_s\gamma$ and $\bar{B}\to X_d\gamma$ in addition to \epsk. Furthermore, such large trilinear couplings can make the EW vacuum unstable. Although the vacuum instability was overlooked in Ref.~\cite{Tanimoto:2016yfy}, we investigate the vacuum (meta-)stability condition in detail and show that the constraint is significant. In Ref.~\cite{Endo:2016aws}, the vacuum condition has been studied in another scenario with large off-diagonal trilinear couplings of the up-type squarks, which bring chargino contributions to the $Z$ penguin. An alternative scenario for the explanation of the \epoe\,discrepancy in the MSSM has been proposed in Ref.~\cite{Kitahara:2016otd,Crivellin:2017gks}. 

The discrepancy in \epoe\,requires large \CPV phases in the off-diagonal trilinear couplings. They also contribute to the branching ratios of $K^+\to\pi^+\nu\bar\nu$ and $K_L\to\pi^0\nu\bar\nu$, the effective branching ratio of $K_S\to \mu^+\mu^-$~\cite{DAmbrosio:2017klp,Chobanova:2017rkj} and the \CP asymmetry difference $\Delta A_{\mathrm{CP}}(b\to s\gamma)$. We investigate SUSY effects on these observables in our scenario, and examine if the effects can be observed at current and/or near-future experiments. 

This paper is organized as follows. In Section \ref{sec:effL} we summarize the effective Lagrangian  together with the RG equations and the one-loop matching conditions that are relevant to our analysis. Top-Yukawa contributions are also explained. In Section \ref{sec:SUSY} we present the gluino contributions associated with the $Z$ penguin. In Section \ref{sec:obs} we explain how each FCNC observable receive gluino contributions. In Section \ref{sec:vacuum} we discuss the constraints from the vacuum stability condition. In Section \ref{sec:analysis} we present our numerical analysis. Our conclusions are drawn in Section~\ref{sec:conclusion}.

\section{Effective Lagrangian and top-Yukawa contributions}
\label{sec:effL}

In this paper, we study flavor-changing processes via the gluino one-loop contributions and the $Z$-boson exchanges.
The latter is described by higher dimensional operators in the SM effective field theory (SMEFT), where the gauge invariance is guaranteed.
The effective Lagrangian is defined as
\begin{align}
  \mathcal{L}_{\rm eff} = \mathcal{L}_{\rm SM} + \sum_i \mathcal{C}_i \mathcal{O}_i,
\end{align}
where the first term in the right-hand side is the SM Lagrangian, and the second one is composed by higher dimensional operators~\cite{Grzadkowski:2010es}. 
In particular, those relevant to the $\Delta F=1$ $Z$-boson penguin are given by 
\begin{align}
  [\mathcal{O}_{HQ}^{(1)}]_{ij} &= (H^\dagger i\overleftrightarrow{D_\mu} H) (\overline{q}_{i} \gamma^\mu q_{j}), \\
  [\mathcal{O}_{HQ}^{(3)}]_{ij} &= (H^\dagger i\overleftrightarrow{D^a_\mu} H) (\overline{q}_{i} \tau^a \gamma^\mu q_{j}), \\
  [\mathcal{O}_{HD}]_{ij} &= (H^\dagger i\overleftrightarrow{D_\mu} H) (\overline{d}_{i} \gamma^\mu d_{j}).
\end{align}
Here, $q$ is the (left-handed) SU(2) quark doublets and $d$ is the (right-handed) down-type quark singlets with quark-flavor indices, $i,j$, and an SU(2) index, $a$.
The Higgs doublet carries a hypercharge $+1/2$, and thus, has a vacuum-expectation value (VEV), $\langle H \rangle = (0,v/\sqrt{2})^T$, with $v \simeq 246\,\GeV$ after the EW symmetry breaking (EWSB).
The covariant derivative is defined for the Higgs doublet as 
\beq
D_{\mu}   =  \partial_{\mu} +i g_2   \frac{\tau^a}{2} W^a_{\mu}   +  i \frac{g_Y}{2}   B_{\mu},
\label{eq:covariantdel}
\eeq
and
\beq
H^\dagger\overleftrightarrow{D^a_\mu} H 
\equiv
H^{\dagger} \tau^a D_{\mu} H - \left( D_{\mu} H\right)^{\dagger}  \tau^a H.
\eeq
On the other hand, $\Delta F=2$ processes are described by the following four-Fermi operators,
\begin{align}
  [\mathcal{O}_{QQ}^{(1)}]_{ijkl} &= (\overline{q}_{i} \gamma_\mu q_{j})(\overline{q}_{k} \gamma^\mu q_{l}), \\
  [\mathcal{O}_{QQ}^{(3)}]_{ijkl} &= (\overline{q}_{i} \tau^a \gamma_\mu q_{j})(\overline{q}_{k} \tau^a \gamma^\mu q_{l}), \\
  [\mathcal{O}_{DD}]_{ijkl} &= (\overline{d}_{i} \gamma_\mu d_{j})(\overline{d}_{k} \gamma^\mu d_{l}), \\
  [\mathcal{O}_{QD}^{(1)}]_{ijkl} &= (\overline{q}_{i} \gamma_\mu q_{j})(\overline{d}_{k} \gamma^\mu d_{l}), \\
  [\mathcal{O}_{QD}^{(8)}]_{ijkl} &= (\overline{q}_{i} \gamma_\mu T_A q_{j})(\overline{d}_{k} \gamma^\mu T^A d_{l}).
\end{align}

The Wilson coefficients develop from the SUSY scale down to the EW one. 
Let us define their beta functions as
\begin{align}
  b_i = (4\pi)^2 \frac{d\,\mathcal{C}_i}{d \ln\mu}.
\end{align}
For the $\mathcal{O}_{HQ}$ and $\mathcal{O}_{HD}$ operators, the relevant terms are~(cf.,~Refs.~\cite{Jenkins:2013zja, Jenkins:2013wua,   Alonso:2013hga})
\begin{align}
  [b_{HQ}^{(1)}]_{12} &= 6 Y_t^2 [\mathcal{C}_{HQ}^{(1)}]_{12}, \notag \\
  [b_{HQ}^{(3)}]_{12} &= 6 Y_t^2 [\mathcal{C}_{HQ}^{(3)}]_{12}, \label{eq:Yt1} \\
  [b_{HD}]_{12} &= 6 Y_t^2 [\mathcal{C}_{HD}]_{12}, \notag
\end{align}
where $Y_t$ is the top-quark Yukawa coupling. 
It is noticed that there are no $\mathcal{O}(\alpha_s)$ corrections at the one-loop level.
The operators also contribute to $\Delta S=2$ four-quark operators as
\begin{align}
  [b_{QQ}^{(1)}]_{1212} &=  \lambda_t Y_t^2 [\mathcal{C}_{HQ}^{(1)}]_{12} + \dots, \notag \\
  [b_{QQ}^{(3)}]_{1212} &= -\lambda_t Y_t^2 [\mathcal{C}_{HQ}^{(3)}]_{12} + \dots, \label{eq:Yt2} \\
  [b_{QD}^{(1)}]_{1212} &=  \lambda_t Y_t^2 [\mathcal{C}_{HD}]_{12} + \dots, \notag
\end{align}
where $[\lambda_t]_{ij} = V_{ti}^*V_{tj}$ and $\lambda_t = [\lambda_t]_{12}$.
In the first leading logarithm approximation, the Wilson coefficients after the RG running from $\Lambda$ to $\mu$ ($\Lambda > \mu$) are estimated as
\begin{align}
  \mathcal{C}_i(\mu) = \mathcal{C}_i(\Lambda) - \frac{1}{(4\pi)^2} b_i(\Lambda) \ln\frac{\Lambda}{\mu}.
    \label{eq:Yt3}
\end{align}
Irrelevant operator mixings and higher-order corrections during the evolutions are neglected. In particular, $\mathcal{C}_{QQ}^{(1)}$, $\mathcal{C}_{QQ}^{(3)}$ and $\mathcal{C}_{QD}^{(1)}$ are generated by $\mathcal{C}_{HQ}$ and $\mathcal{C}_{HD}$.

After the EWSB, $\mathcal{O}_{HQ}$ and $\mathcal{O}_{HD}$ are matched to the flavor-changing $Z$ couplings through the expansion,
\begin{align}
 H^\dagger i\overleftrightarrow{D_\mu} H &= 
 \frac{g_Z}{2}v^2 Z_{\mu}
 + G^- i \overleftrightarrow{\partial_\mu}   G^+
 - g_2 v \left( W^+_{\mu} G^- + W^-_{\mu} G^+ \right) +  \dots,\\
 H^\dagger i\overleftrightarrow{D_\mu^3} H &=  
 - \frac{g_Z}{2}v^2 Z_{\mu}
 + G^- i \overleftrightarrow{\partial_\mu}   G^+
 + \dots
\end{align}
with $g_Z = \sqrt{g_2^2  + g_Y^2}$, where the terms irrelevant for the matching onto the $\Delta S=2$ operators are omitted.

The operators also contribute to $\Delta F=2$ observables through the effective Hamiltonian,
\begin{align}
 \mathcal{H}_{\rm eff} = 
 \sum_{i=1}^5 \mathcal{C}_i \mathcal{O}_i + \sum_{i=1}^3  \mathcal{C}'_i \mathcal{O}'_i + \textrm{H.c.},
\end{align}
where the effective operators are
\begin{align}
 [\mathcal{O}_1]_{ij} &= 
 (\bar d_i^\alpha \gamma_\mu P_L d_j^\alpha)(\bar d_i^\beta \gamma^\mu P_L d_j^\beta),\\
 [\mathcal{O}_2]_{ij} &= 
 (\bar d_i^\alpha P_L d_j^\alpha)(\bar d_i^\beta P_L d_j^\beta),\\
 [\mathcal{O}_3]_{ij} &= 
 (\bar d_i^\alpha P_L d_j^\beta)(\bar d_i^\beta P_L d_j^\alpha),\\
 [\mathcal{O}_4]_{ij} &= 
 (\bar d_i^\alpha P_L d_j^\alpha)(\bar d_i^\beta P_R d_j^\beta),\\
 [\mathcal{O}_5]_{ij} &= 
 (\bar d_i^\alpha P_L d_j^\beta)(\bar d_i^\beta P_R d_j^\alpha),
\end{align}
with color indices $\alpha, \beta$.
In this paper, chirality-flipped operators and their Wilson coefficients are denoted with a prime. 
At the tree level, the SMEFT operators are matched at the weak scale to these operators as \cite{Aebischer:2015fzz} 
\begin{align}
 [\mathcal{C}_1]_{ij}^{(0)} &= - \left( [\mathcal{C}_{QQ}^{(1)}]_{ijij} + [\mathcal{C}_{QQ}^{(3)}]_{ijij} \right),~~~
 [\mathcal{C}'_1]_{ij}^{(0)} = - [\mathcal{C}_{DD}]_{ijij}, \label{eq:SMEFT4Q1} \\
 [\mathcal{C}_4]_{ij}^{(0)} &= [\mathcal{C}_{QD}^{(8)}]_{ijij}, \label{eq:SMEFT4Q2} \\
 [\mathcal{C}_5]_{ij}^{(0)} &= 2[\mathcal{C}_{QD}^{(1)}]_{ijij} - \frac{1}{N_c} [\mathcal{C}_{QD}^{(8)}]_{ijij}, \label{eq:SMEFT4Q3} 
\end{align}
where $N_c=3$ is the number of colors.
In addition, these low-energy $\Delta F = 2$ operators are generated by the $\Delta F = 1$ ones in the SMEFT through the one-loop matchings at the weak scale~\cite{Aebischer:2015fzz}.  
The conditions for $\mathcal{C}_{HQ}$ and $\mathcal{C}_{HD}$ at the scale $\mu_W$ are approximated as~\cite{Endo:2016tnu, Bobeth:2017xry}
\begin{align}
 [\mathcal{C}_1]_{ij}^{(1)} &= \frac{\alpha [\lambda_t]_{ij}}{\pi s_W^2}
 \left[ 
 [\mathcal{C}_{HQ}^{(1)}]_{ij}\,I_1(x_t,\mu_W) 
 - [\mathcal{C}_{HQ}^{(3)}]_{ij}\,I_2(x_t,\mu_W) 
 \right], 
 \label{eq:matching1} \\
 [\mathcal{C}_5]_{ij}^{(1)} &= -\frac{2\alpha [\lambda_t]_{ij}}{\pi s_W^2} 
 [\mathcal{C}_{HD}]_{ij}\,  I_1(x_t,\mu_W),
 \label{eq:matching2}
\end{align}
with $x_t = m_t^2/m_W^2$. 
These results are gauge-independent.
The loop functions are defined as
\begin{align}
  I_1(x,\mu) &= \frac{x}{8} \left[ 
  \ln\frac{\mu}{m_W} - \frac{x-7}{4(x-1)} - \frac{x^2-2x+4}{2(x-1)^2}\ln x
  \right], \\
  I_2(x,\mu) &= \frac{x}{8} \left[ 
  \ln\frac{\mu}{m_W} + \frac{7x-25}{4(x-1)} - \frac{x^2-14x+4}{2(x-1)^2}\ln x
  \right].
\end{align}
Here, we discarded box contributions which are suppressed by CKM factors or by $m_{c,u}^2/m_W^2$ in the $\Delta S = 2$ case (see Ref.~\cite{Bobeth:2017xry}).

The RG equations in Eqs.~\eqref{eq:Yt1} and \eqref{eq:Yt2} and the matching conditions in Eqs.~\eqref{eq:matching1} and \eqref{eq:matching2} are proportional to $Y_t^2$, and hence, we call them the top-Yukawa contributions.

\section{SUSY contributions}
\label{sec:SUSY}

\begin{figure}[t]
\begin{center}
\subfigure[]{
\raisebox{9.5mm}{\includegraphics[width=0.2\textwidth, bb= 0 0 137 91]{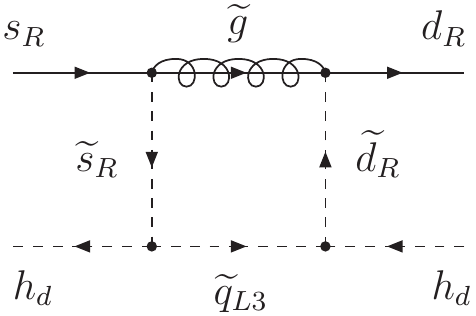}}
}
\subfigure[]{
\includegraphics[width=0.2\textwidth, bb= 0 0 137 130]{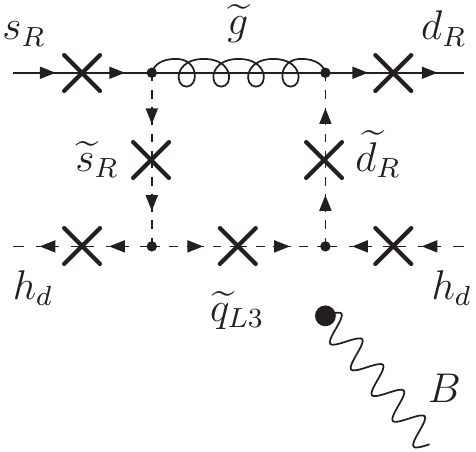}
}
\subfigure[]{
\includegraphics[width=0.2\textwidth, bb= 0 0 137 130]{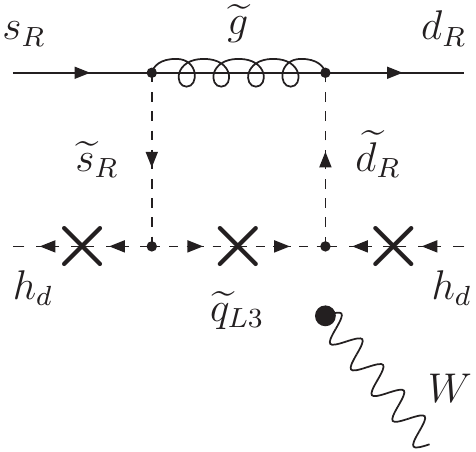}
}
\subfigure[]{
\includegraphics[width=0.2\textwidth, bb= 0 0 137 130]{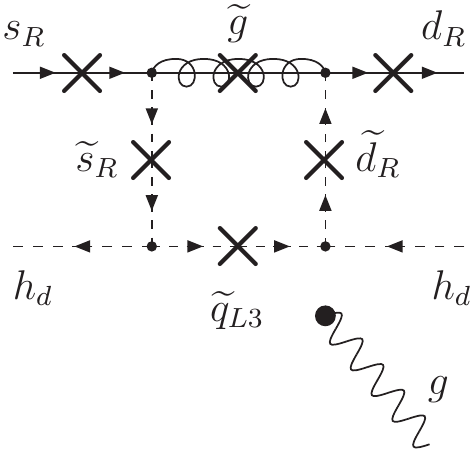}
}
\subfigure[]{
\raisebox{3mm}{\includegraphics[width=0.2\textwidth, bb= 0 0 137 83]{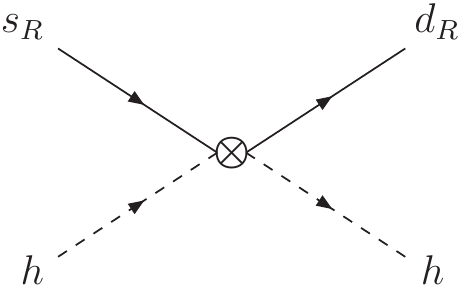}}
}
\subfigure[]{
\includegraphics[width=0.2\textwidth, bb= 0 0 133 95]{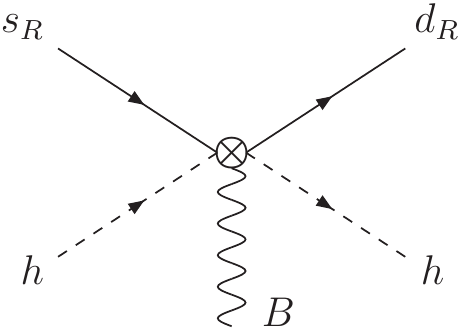}
}
\subfigure[]{
\includegraphics[width=0.2\textwidth, bb= 0 0 131 110]{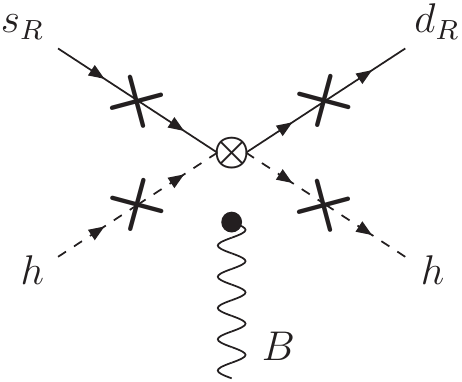}
}
\subfigure[]{
\includegraphics[width=0.2\textwidth, bb= 0 0 133 95]{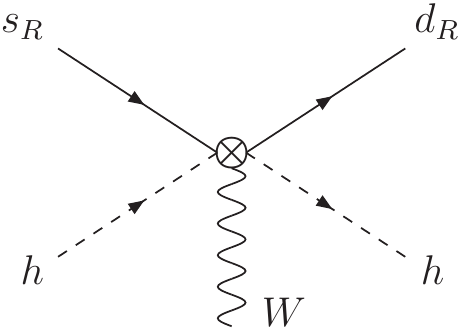}
}
\subfigure[]{
\includegraphics[width=0.2\textwidth, bb= 0 0 133 110]{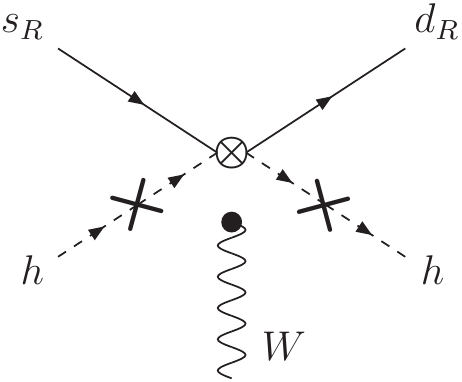}
}
\subfigure[]{
\includegraphics[width=0.2\textwidth, bb= 0 0 133 110]{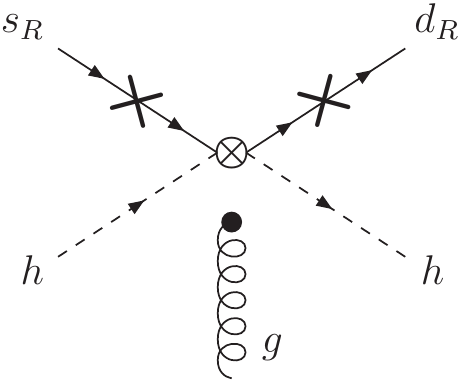}
}
\caption{
Feynman diagrams relevant for the matchings onto the operators $[\mathcal{O}_{HD}]_{12}$, where the external gauge bosons are attached to each of the cross marks.
Diagrams (a)--(d) are the one-loop gluino contributions, and (e)--(j) are the diagrams in the SMEFT.
The diagrams contributing to $[\mathcal{O}_{HQ}^{(1,3)}]_{12}$ are similarly obtained.
}
\label{fig:diagrams}
\end{center}
\end{figure}

At the one-loop level, $\mathcal{O}_{HQ}$ and $\mathcal{O}_{HD}$ are generated by gluino loops in the MSSM.
When the squark (quark) flavor is violated by scalar trilinear soft-breaking parameters, the dominant contributions are calculated from Fig.~\ref{fig:diagrams} as
\begin{align}
  [\mathcal{C}_{HQ}^{(1)}]_{12} &= -\frac{\alpha_s}{12\pi}\frac{\cos^2\beta}{m_{\tilde g}^4} (T_D)_{13}^* (T_D)_{23}\,
  Z(x_{L1}, x_{L2}, x_{R3}), \label{eq:CHQ}\\
  [\mathcal{C}_{HQ}^{(3)}]_{12} &= -\frac{\alpha_s}{12\pi}\frac{\cos^2\beta}{m_{\tilde g}^4} (T_D)_{13}^* (T_D)_{23}\,
  Z(x_{L1}, x_{L2}, x_{R3}),\label{eq:CHQ3} \\
  [\mathcal{C}_{HD}]_{12} &= \frac{\alpha_s}{6\pi}\frac{\cos^2\beta}{m_{\tilde g}^4} (T_D)_{31} (T_D)_{32}^*\,
  Z(x_{R1}, x_{R2}, x_{L3}), \label{eq:CHD}
\end{align}
with $x_i=m_{\tilde d_i}^2/m_{\tilde g}^2$. 
Here, $m_{\tilde d_{L(R)i}}$ is the left- (right-) handed squark soft mass for the $i$-th generation, $m_{\tilde g}$ is the gluino mass, and $T_D$ is the scalar trilinear coupling of the down-type squarks.
In this paper, the SUSY Les Houches Accord (SLHA) notation~\cite{Skands:2003cj, Allanach:2008qq} is used, and flavor violations are discussed in the basis where the Yukawa matrix of the down-type quark is diagonalized. 
The Wilson coefficients are set at the SUSY scale.\footnote{
 If the trilinear couplings $(T_D)_{13,23,31,32}$ are set in a scale higher than the SUSY scale, the flavor-violating squark soft masses $(m_{\tilde d_{L(R)}})_{12,21}$ are generated via RG corrections. 
They can be sizable and contribute to the kaon FCNCs when the input scale is much higher than the SUSY scale.
}
The loop function is defined as
\begin{align}
 Z(x,y,z) =&~ -\frac{x^2\ln x}{(x-1)(x-y)(x-z)^2} + \frac{y^2\ln y}{(y-1)(x-y)(y-z)^2} \notag\\
 &~ - \frac{z}{(z-1)(x-z)(y-z)} + \frac{(2xy-yz-xz-xyz+z^3)z\ln z}{(z-1)^2(x-z)^2(y-z)^2}.
\end{align}
In the limit of $y,z \to x$, it becomes
\begin{align}
 Z(x) = \frac{2+3x-6x^2+x^3+6x\ln x}{6x(x-1)^4}.
\end{align}
Other SUSY contributions are explained in the next section. 

Note that, in literature, e.g., Ref.~\cite{Bertolini:1990if}, it has been argued that gluino-mediated contributions to EW penguin are suppressed compared to the other penguins, by assuming that the gluino contributions to the EW penguin are proportional to those to the photon penguin.
However, this is not the case in our scenario, where the SU(2$)\times$U(1) symmetry is broken by large scalar trilinear couplings. 
Such couplings can generate the $Z$ penguin significantly via double-mass insertion contributions, as was pointed out in Ref.~\cite{Colangelo:1998pm} and explicitly shown in this section.

\section{Observables}
\label{sec:obs}

\subsection{$\boldsymbol{\epoe}$}

The direct \CP violation of the $K \to \pi \pi $ decays, \epoe, includes the SM and SUSY $Z$-penguin contributions,
\begin{align}
 \left( \epoe \right) = 
 \left( \epoe \right)^{\rm SM} +
 \left( \epoe \right)^{\rm SUSY}.
\end{align}
The latter contribution is approximated to be (cf., Ref.~\cite{Buras:2015jaq})\footnote{
  Another SUSY contribution is produced from chromomagnetic-dipole diagrams \cite{Masiero:1999ub, Babu:1999xf, Khalil:1999ym, Baek:1999jq, Barbieri:1999ax, Buras:1999da, Baek:2001kc, Chobanova:2017rkj}. 
  The Wilson coefficient is obtained by replacing $b \to s$ and $d_{i} \to d$ in Eq.~\eqref{eq:gluinochromo}.
  In our analyses, such a contribution is negligible because the squark mixings between the fist two generations are assumed to be suppressed. 
}
\begin{align}
 \left( \epoe \right)^{\rm SUSY} &= 
  - {B_8^{(3/2)}(m_c)}\bigg[
 5.91 \times 10^7\GeV^2\,{\rm Im}
 \left([\mathcal{C}_{HQ}^{(1)}]_{12}+[\mathcal{C}_{HQ}^{(3)}]_{12}\right)
 \notag \\ &\qquad\qquad\qquad~~~
 + 1.97 \times 10^8\GeV^2\,{\rm Im}\,[\mathcal{C}_{HD}]_{12}
 \bigg],
 \label{eq:epsPrime}
\end{align}
where the Wilson coefficients are estimated at the $Z$-boson mass scale, $\mu=m_Z$. 
By using lattice simulations~\cite{Blum:2011ng, Blum:2012uk, Blum:2015ywa}, $B_8^{(3/2)}(m_c) = 0.76 \pm 0.05$ is obtained~\cite{Buras:2015yba, Buras:2015qea}.
Here, \epsk\,in the denominator is evaluated by the experimental value.
The right-handed contribution is amplified by $c_W^2/s_W^2 \simeq 3.33$ compared to the left-handed one.

Currently, the SM prediction deviates from the experimental result at the $2.8\,\sigma$ level.
In this paper, the discrepancy of \epoe\,is required to be explained within the $1\,\sigma$ range,
\begin{align}
 10.0\times 10^{-4} < \left( \epoe \right)^{\rm SUSY} < 21.1 \times 10^{-4},
 \label{eq:eps_limit}
\end{align}
where Ref.~\cite{Kitahara:2016nld} is used for the SM prediction at the NLO level.

\subsection{$\boldsymbol{\epsk}$}

Both the SM and SUSY affect to the indirect \CP violation of the neutral kaon system,
\beq
\epsk = e^{i \varphi_{\varepsilon}} \left( \epsk^{\rm SM} + \epsk^{\rm SUSY} \right),
\eeq
where $\varphi_{\varepsilon} = ( 43.51\pm  0.05 )^{\circ}$.
$\epsk^{\rm SUSY}$ is composed by gluino box diagrams as well as $\mathcal{C}_{HQ}$ and $\mathcal{C}_{HD}$. 
In our scenario, although the gluino box contributions are sizable, their dominant contributions arise as dimension-ten operators in the SMEFT.
In order to include them in our formalism, we separately calculate them in the broken phase, where the Higgs VEV is involved.\footnote{
 Equations~\eqref{eq:SMEFT4Q1}--\eqref{eq:SMEFT4Q3} are not used for evaluating the gluino box contributions to the $\Delta S=2$ observables. 
}
At the one-loop level, they are obtained as~\cite{Hagelin:1992tc}
\begin{align}
 [\mathcal{C}_1]_{ij} &= \frac{\alpha_s^2}{m_{\tilde g}^2}
 \mathcal{R}_{ri}^{d*}\mathcal{R}_{rj}^{d} \mathcal{R}_{si}^{d*}\mathcal{R}_{sj}^{d}
 \left[ \frac{1}{9} B_0(x_r,x_s) + \frac{11}{36} B_2(x_r,x_s) \right], \\
 [\mathcal{C}_2]_{ij} &= \frac{\alpha_s^2}{m_{\tilde g}^2}
 \mathcal{R}_{r,i+3}^{d*}\mathcal{R}_{rj}^{d} \mathcal{R}_{s,i+3}^{d*}\mathcal{R}_{sj}^{d}
 \left[ \frac{17}{18} B_0(x_r,x_s) \right], \\
 [\mathcal{C}_3]_{ij} &= \frac{\alpha_s^2}{m_{\tilde g}^2}
 \mathcal{R}_{r,i+3}^{d*}\mathcal{R}_{rj}^{d} \mathcal{R}_{s,i+3}^{d*}\mathcal{R}_{sj}^{d}
 \left[ - \frac{1}{6} B_0(x_r,x_s) \right], \\
 [\mathcal{C}_4]_{ij} &= \frac{\alpha_s^2}{m_{\tilde g}^2}\Bigg\{
 \mathcal{R}_{ri}^{d*}\mathcal{R}_{rj}^{d} \mathcal{R}_{s,i+3}^{d*}\mathcal{R}_{s,j+3}^{d}
 \left[ \frac{7}{3} B_0(x_r,x_s) - \frac{1}{3} B_2(x_r,x_s)
 \right] \notag \\
 &~~~~~~~~~~ + 
 \mathcal{R}_{ri}^{d*}\mathcal{R}_{r,j+3}^{d} \mathcal{R}_{s,i+3}^{d*}\mathcal{R}_{sj}^{d}
 \left[ - \frac{11}{18} B_2(x_r,x_s) \right] \Bigg\}, \\
 [\mathcal{C}_5]_{ij} &= \frac{\alpha_s^2}{m_{\tilde g}^2}\Bigg\{
 \mathcal{R}_{ri}^{d*}\mathcal{R}_{rj}^{d} \mathcal{R}_{s,i+3}^{d*}\mathcal{R}_{s,j+3}^{d}
 \left[ \frac{1}{9} B_0(x_r,x_s) + \frac{5}{9} B_2(x_r,x_s)
 \right] \notag \\
 &~~~~~~~~~~ + 
 \mathcal{R}_{ri}^{d*}\mathcal{R}_{r,j+3}^{d} \mathcal{R}_{s,i+3}^{d*}\mathcal{R}_{sj}^{d} 
 \left[ -\frac{5}{6} B_2(x_r,x_s) \right] \Bigg\},
\end{align}
at the SUSY scale ($\mu_{\textrm{SUSY}}$) with generation indices $i \neq j$ and $x_r = m_{\tilde d_r}^2/m_{\tilde g}^2$, where $\mathcal{R}^d_{r i }$ for $r = 1,2,\ldots,6$ is the squark rotation matrix defined in the SLHA notation~\cite{Skands:2003cj, Allanach:2008qq}.
$\mathcal{C}^{\prime}_{1,2,3}$ are obtained by flipping the chirality of $\mathcal{R}_{ri}^{d (*)}$ in $\mathcal{C}_{1,2,3}$.
The loop functions are defined as
\begin{align}
 B_0(x,y) &= \frac{x \ln x}{(x-y)(x-1)^2}+\frac{y \ln y}{(y-x)(y-1)^2}+\frac{1}{(x-1)(y-1)}, \\
 B_2(x,y) &= \frac{x^2 \ln x}{(x-y)(x-1)^2}+\frac{y^2 \ln y}{(y-x)(y-1)^2}+\frac{1}{(x-1)(y-1)}.
\end{align}
From $\mu_{\textrm{SUSY}}$ to the hadronic scale, we solve the RG equations at the NLO level \cite{Buras:2001ra} and use the hadronic matrix elements in Ref.~\cite{Garron:2016mva}.

Additionally, $[\mathcal{C}_1]_{ij}$ and $[\mathcal{C}_5]_{ij}$ receive the top-Yukawa contributions depending on $\mathcal{C}_{HQ}$ and $\mathcal{C}_{HD}$ as
\begin{align}
 [\mathcal{C}_1]_{ij} &= \frac{\alpha [\lambda_t]_{ij}}{\pi s_W^2}
 \left[ 
 [\mathcal{C}_{HQ}^{(1)}]_{ij}\,I_1(x_t,\mu_{\rm SUSY}) 
 - [\mathcal{C}_{HQ}^{(3)}]_{ij}\,I_2(x_t,\mu_{\rm SUSY}) 
 \right], \label{eq:epsKtop1} \\
 [\mathcal{C}_5]_{ij} &= -\frac{2\alpha [\lambda_t]_{ij}}{\pi s_W^2} 
 [\mathcal{C}_{HD}]_{ij}\,  I_1(x_t,\mu_{\rm SUSY}),
 \label{eq:epsKtop2}
\end{align}
at the $Z$-boson mass scale.
These results are derived as follows: 
The Wilson coefficients are evolved by solving the RG equations with the beta function \eqref{eq:Yt2} in the first leading logarithm approximation \eqref{eq:Yt3}, and then, matched onto the low-scale operators at the weak scale \eqref{eq:SMEFT4Q1}--\eqref{eq:SMEFT4Q3}.
Also, the one-loop matchings, \eqref{eq:matching1} and \eqref{eq:matching2}, are taken into account to include the additional contributions of $\mathcal{C}_{HQ}$ and $\mathcal{C}_{HD}$ at the weak scale (see  Ref.~\cite{Bobeth:2017xry}).\footnote{
 The results are independent of the matching scale $\mu_W$ by including the one-loop matching conditions.
 Consequently, the logarithmic function becomes $\ln(\mu_{\rm SUSY}/m_W)$.
}
Equivalently, the same results are reproduced by substituting $\mu_W \to \mu_{\rm SUSY}$ in Eqs.~\eqref{eq:matching1} and \eqref{eq:matching2}.
This is because the logarithmic scale dependence of the one-loop matching conditions has the same origin as the one-loop beta functions (see Ref.~\cite{Endo:2016tnu}).

It is also noticed that, in Eq.~\eqref{eq:epsKtop1}, the logarithmic dependence of $\mu_{\rm SUSY}$ cancels out because of $[\mathcal{C}_{HQ}^{(1)}]_{12} = [\mathcal{C}_{HQ}^{(3)}]_{12}$ in Eqs.~\eqref{eq:CHQ} and \eqref{eq:CHQ3}.
On the other hand, the scale dependence in Eq.~\eqref{eq:epsKtop2} remains, and thus, $[\mathcal{C}_5]_{ij}$ is sensitive to $\mu_{\rm SUSY}$.

The SM value is estimated to be
\begin{align}
 \epsk^{\rm SM} = (2.12 \pm 0.18) \times 10^{-3},
 \label{eq:epsKSM}
\end{align}
where the input SM parameters are found in Ref.~\cite{Jang:2017ieg} (cf., Ref.~\cite{Bailey:2015tba}).
Especially, the Wolfenstein parameters are determined by the angle-only fit~\cite{Bevan:2013kaa}, and $|V_{cb}|$ obtained from inclusive semileptonic $B$ decays $(\bar{B} \to X_c \ell^{-} \bar{\nu})$ \cite{Amhis:2016xyh} is used.\footnote{
 Recently, there are debates about systematic uncertainties of the exclusive determinations of $|V_{cb}|$
 ~\cite{Bigi:2017njr,Grinstein:2017nlq,Bernlochner:2017xyx}.
}
We use lattice results for the $\xi_0$ parameter~\cite{Bai:2015nea}, which parametrizes the absorptive part of long-distance effects, and refrain from relying on the experimental result of \epoe, because we consider SUSY contributions to \epoe.
On the other hand, the experimental result is~(cf., Ref.~\cite{Olive:2016xmw})
\begin{align}
 |\epsk^{\rm exp}| = (2.228 \pm 0.011) \times 10^{-3}.
\end{align}
Therefore, the SUSY contributions are required to be within the range,
\begin{align}
 -0.25 \times 10^{-3} < \epsk^{\rm SUSY} < 0.47 \times 10^{-3},
\end{align}
at the $2\,\sigma$ level.\footnote{
In our analysis, the gluino contributions are much less constrained by the mass difference of the neutral kaons, $\Delta M_K$, because hadronic uncertainties are large.
}

\subsection{$\boldsymbol{K \to \pi\nu\bar\nu}$}

The $Z$-penguin contributions induce the decays, $K^+ \to \pi^+\nu\bar\nu$ and $K_L \to \pi^0\nu\bar\nu$. 
They are expressed as
\cite{Buras:2015jaq, Buras:2015qea}
\begin{align}
 \mathcal{B}(K^+\to\pi^+\nu\bar\nu) &= 
 \kappa_+ \left[
 \left(
 \frac{{\rm Im}\,X_{\rm eff}}{\lambda^5} 
 \right)^2 + 
 \left(
 \frac{{\rm Re}\,\lambda_c}{\lambda}P_c(X) + 
 \frac{{\rm Re}\,X_{\rm eff}}{\lambda^5} 
 \right)^2
 \right], \\
 \mathcal{B}(K_L\to\pi^0\nu\bar\nu) &= 
 \kappa_L \left[
 \frac{{\rm Im}\,X_{\rm eff}}{\lambda^5} 
 \right]^2,
\end{align}
where $\lambda = |V_{us}|$, $\lambda_c = V_{cd}^*V_{cs}$, $\kappa_+ = (5.157 \pm 0.025) \times 10^{-11}(\lambda/0.225)^8$, $\kappa_L = (2.231 \pm 0.013) \times 10^{-10}(\lambda/0.225)^8$, and the charm contribution gives $P_c(X)= (9.39 \pm 0.31)\times 10^{-4} /\lambda^4  + (0.04 \pm 0.02)$. 
In terms of $\mathcal{C}_{HQ}$ and $\mathcal{C}_{HD}$, $X_{\rm eff}$ is approximated to be (cf., Ref.~\cite{Endo:2016tnu})
\begin{align}
 {\rm Re}\,X_{\rm eff} &= -4.83\times10^{-4}
 -5.62\times10^6\GeV^2\,
 {\rm Re}\,\mathcal{C}_{H+}, \\
 {\rm Im}\,X_{\rm eff} &= 2.12\times10^{-4}
 +5.62\times10^6\GeV^2\,
 {\rm Im}\,\mathcal{C}_{H+},
 \label{eq:KLpinn}
\end{align}
where the first terms in the right-hand sides are the SM contributions in each equation, and
\begin{align}
 \mathcal{C}_{H+} = [\mathcal{C}_{HQ}^{(1)}]_{12}+[\mathcal{C}_{HQ}^{(3)}]_{12}+[\mathcal{C}_{HD}]_{12}.
\end{align}
The Wilson coefficients are estimated at the $Z$-boson mass scale.

The SM predictions are known to be~\cite{Endo:2016tnu}
\begin{align}
 \mathcal{B}(K^+\to\pi^+\nu\bar\nu)^{\rm SM} &= (8.5 \pm 0.5) \times 10^{-11}, \\
 \mathcal{B}(K_L\to\pi^0\nu\bar\nu)^{\rm SM} &= (3.0 \pm 0.2) \times 10^{-11},
\end{align}
while the experimental results are \cite{Artamonov:2008qb, Ahn:2009gb}
\begin{align}
 \mathcal{B}(K^+\to\pi^+\nu\bar\nu)^{\rm exp} &= (17.3^{+11.5}_{-10.5}) \times 10^{-11}, \\
 \mathcal{B}(K_L\to\pi^0\nu\bar\nu)^{\rm exp} & < 2.6 \times 10^{-8}.~~~[90\%~\mbox{C.L.}]
\end{align}
These experimental values will be improved in the near future.
The NA62 experiment at CERN has already started the physics run 
and aims to measure $ \mathcal{B}(K^+\to\pi^+\nu\bar\nu)$ with a precision of $10 \%$ relative to the SM prediction~\cite{NA62:2017rwk}.
The KOTO experiment at J-PARC aims to measure $\mathcal{B}(K_L\to\pi^0\nu\bar\nu)$ 
around the SM sensitivity by 2021~\cite{KOTOfuture1,KOTOfuture2}.

\subsection{$\boldsymbol{K_L \to \mu^+\mu^-}$}

The decay rate of $K_L \to \mu^+\mu^-$, which is  a \CPC process, is sensitive to a real component of the flavor-changing $Z$ couplings. 
There are large theoretical uncertainties from a long-distance (LD) contribution.
In addition, an unknown sign of $\mathcal{A}\left( K_L \to \gamma \gamma  \right)$   conceals a relative sign between the LD and a short-distance (SD) amplitudes.
One can, therefore, estimate only the SD branching ratio, which is expressed as~\cite{Buras:2015jaq,Gorbahn:2006bm,Bobeth:2013tba}
\begin{align}
 \mathcal{B}(K_L \to \mu^+\mu^-)_{\rm SD} &= 
 \kappa_\mu 
 \left(
 \frac{{\rm Re}\,\lambda_c}{\lambda}P_c(Y) + 
 \frac{{\rm Re}\,Y_{\rm eff}}{\lambda^5} 
 \right)^2,
\end{align}
where $\kappa_\mu=(2.01\pm0.02)\times 10^{-9}(\lambda/0.225)^8$, and the charm-quark contribution is $P_c(Y)= (0.115 \pm 0.018)\times (0.225/\lambda)^4$. 
Here, $Y_{\rm eff}$ is approximately given as~(cf., Ref.~\cite{Endo:2016tnu})
\begin{align}
 {\rm Re}\,Y_{\rm eff} &= -3.07\times10^{-4}-
 5.62\times10^6\GeV^2\,{\rm Re}\,\mathcal{C}_{H-},
\end{align}
where the first term in the right-hand side is the SM contribution, and
\begin{align}
 \mathcal{C}_{H-} = [\mathcal{C}_{HQ}^{(1)}]_{12}+[\mathcal{C}_{HQ}^{(3)}]_{12}-[\mathcal{C}_{HD}]_{12} .
\end{align}
The Wilson coefficients are estimated at the $Z$-boson mass scale.

The SM value is obtained as~\cite{Endo:2016tnu}
\begin{align}
 \mathcal{B}(K_L \to \mu^+\mu^-)_{\rm SD}^{\rm SM} = (0.83 \pm 0.10) \times 10^{-9}.
\end{align}
It is challenging to extract the SD contribution from the experimental value.
An upper bound is estimated as \cite{Isidori:2003ts}
\begin{align}
 \mathcal{B}(K_L \to \mu^+\mu^-)_{\rm SD}^{\rm exp} < 2.5 \times 10^{-9}.
\end{align}
Since the constraint is much weaker than the SM uncertainties, we simply impose a bound,
\begin{align}
 -1.81 \times 10^{-10}~(\textrm{GeV})^{-2} 
 < {\rm Re}\,\mathcal{C}_{H-} 
 < 4.85 \times 10^{-11} ~(\textrm{GeV})^{-2}.
\end{align}

\subsection{$\boldsymbol{K_S \to \mu^+\mu^-}$}

The decay, $K_S \to \mu^+\mu^-$, proceeds via LD \CPC P-wave and SD \CPV S-wave processes.
Since the decay rate is dominated by the former, whose uncertainty is large, the sensitivity to the imaginary component of the flavor-changing $Z$ couplings is diminished~\cite{Ecker:1991ru,Isidori:2003ts,Mescia:2006jd}.
Interestingly, the SD contribution is enhanced through an interference between the $K_L$ and $K_S$ states in the neutral kaon beam~\cite{DAmbrosio:2017klp}.
The effective branching ratio of $ K_S \to \mu^+ \mu^- $ after including the interference is expressed as (cf., Ref.~\cite{DAmbrosio:2017klp})
\begin{align}
 \mathcal{B} ( K_S \to \mu^+ \mu^- )_{\rm eff} = 
 \mathcal{B} ( K_S \to \mu^+ \mu^- ) 
 + D  \cdot \mathcal{B} ( K_S \to \mu^+ \mu^- )_{\rm int},
\end{align}
where a dilution factor $D$ is an initial asymmetry between the numbers of $K^0$ and $\overline{K}{}^0$,
\begin{align}
 D = \left( K^0 - \overline{K}{}^0 \right) /  \left( K^0 + \overline{K}{}^0 \right).
\end{align}
In the right-hand side, the branching ratio is approximated to be
\begin{align}
 \mathcal{B} ( K_S \to \mu^+ \mu^- ) &= 
 4.99 \times 10^{-12} + 3.30 \times 10^{8}\GeV^4 \left[ 2.39 \times 10^{-11} \GeV^{-2} + {\rm Im}\,\mathcal{C}_{H-}\right]^2,
 \label{eq:KSmmBr} 
\end{align}
where the first and second terms in the right-hand side come from the LD and SD contributions, respectively. 
Here, the Wilson coefficients are estimated at the $Z$-boson mass scale.
On the other hand, the interference contribution is given as
\begin{align}
 \mathcal{B}( K_S \to \mu^+ \mu^- )_{\textrm{int}} =
 \left\{ 
 \begin{array}{ll}
 - 7.69\times10^{7}\GeV^4 \left[ 2.39 \times 10^{-11}\GeV^{-2} + \textrm{Im}\,\mathcal{C}_{H-} \right]
 & \\ \qquad\qquad\qquad~~~~ \times 
 \left[ 1.73\times 10^{-9}\GeV^{-2}  -  \textrm{Re}\,\mathcal{C}_{H-} \right], & (\eta_\mathcal{A}=+) \\
 7.69\times10^{7}\GeV^4 \left[ 2.39 \times 10^{-11}\GeV^{-2}  + \textrm{Im}\,\mathcal{C}_{H-} \right]
 & \\ \qquad\qquad\qquad~~~~ \times
 \left[ 1.86\times 10^{-9}\GeV^{-2}  +  \textrm{Re}\,\mathcal{C}_{H-} \right].  & (\eta_\mathcal{A}=-)
 \end{array}
 \right.
\label{eq:Breff}
\end{align}
The Wilson coefficients are estimated at the $Z$-boson mass scale.
The unknown relative sign between the LD and SD contributions in $K_L \to \mu^+ \mu^-$ gives 
two different predictions of $\mathcal{B} \left( K_S \to \mu^+ \mu^- \right)_{\rm int}$, which are expressed by $\eta_\mathcal{A}$, (see Ref.~\cite{Cirigliano:2011ny,DAmbrosio:2017klp})
\begin{align}
 \eta_\mathcal{A} = \textrm{sgn}
 \left[
 \frac{\mathcal{A}\left( K_L \to \gamma \gamma \right)}{\mathcal{A} \left( K_L \to (\pi^0)^{\ast} \to \gamma \gamma \right)}
 \right].
\end{align}
Here, scalar operator contributions are discarded in the above formulae: they can be significant especially when $\tan \beta$ is large and $m_A$ is small~\cite{Chobanova:2017rkj}.

The SM prediction depends on $D$ and $\eta_\mathcal{A}$, which are determined by experiments.
For $D=0$, it is obtained as~\cite{Ecker:1991ru,Isidori:2003ts,DAmbrosio:2017klp}
\begin{align}
\mathcal{B}(K_S \to \mu^+ \mu^- )^{\rm SM} = \left( 5.18  \pm 1.50 \right) \times 10^{-12},
\label{eq:KSMUMU_SMD0}
\end{align}
while for $D=1$ and $\eta_\mathcal{A}=-1$, the SM prediction becomes~\cite{DAmbrosio:2017klp}
\begin{align}
\mathcal{B}(K_S \to \mu^+ \mu^- )_{\rm eff}^{\rm SM} = \left( 8.59 \pm 1.50 \right) \times 10^{-12}.
\label{eq:KSMUMU_SMD1}
\end{align}
On the other hand, the current experimental bound based on the LHCb Run-1 result using the integrated luminosity 3\,fb${}^{-1}$ is~\cite{Aaij:2017tia}
\begin{align}
\mathcal{B}(K_S \to \mu^+ \mu^- )^{\rm exp} < 0.8 \times 10^{-9}. ~~~[90\%~\mbox{C.L.}]
\end{align}
The experimental sensitivity is expected to reach $\mathcal{B}(K_S \to \mu^+ \mu^- ) = \mathcal{O}(10^{-11})$ by the end of the LHCb Run-2, and the Run-3 project is aiming to achieve the sensitivity as precise as the SM level \cite{LHCbupgrade}.

\subsection{$\boldsymbol{b\to d\gamma}$ and $\boldsymbol{b\to s\gamma}$}

In this paper, we consider flavor-violations in the scalar trilinear couplings.
They contribute to the decays of $b\to d_i\gamma$ $(d_i=d,s)$ at the one-loop level.\footnote{
 They also contribute to the ($CP$-violating) $B_{d,s}$ mixings.
 In the parameter regions of our interest, 
 gluino box contributions to them are smaller than the current experimental and theoretical uncertainties. 
 Also, the $CP$-violating scalar trilinear couplings can contribute to the electric dipole moments (EDMs) e.g., of the neutron. 
 Since the $CP$ phases are introduced in the flavor off-diagonal components, 
 the gluino contributions to the EDMs satisfy the experimental limits.
}
The decays are described by the effective Hamiltonian,
\begin{align}
 \mathcal{H}_{\rm eff} = 
 -\frac{4G_F}{\sqrt{2}} [\lambda_t]_{i3} \Big[ 
 \mathcal{C}_{7\gamma} \mathcal{O}_{7\gamma} + \mathcal{C}_{8g} \mathcal{O}_{8g} \Big] + (L\leftrightarrow R),
\end{align}
where the effective operators are defined as
\begin{align}
 \mathcal{O}_{7\gamma} = \frac{e}{16\pi^2} m_b\, \bar d_i \sigma^{\mu\nu} P_R b\, F_{\mu\nu},~~~
 \mathcal{O}_{8g} = \frac{g_3}{16\pi^2} m_b\, \bar d_i \sigma^{\mu\nu} T^a P_R b\, G_{\mu\nu}^a,
\end{align}
where $e>0$ and $g_3 >0$, and the covariant derivatives for the quark and squark follow the same sign convention as Eq.~\eqref{eq:covariantdel}.
At the one-loop level, the gluino contributions are obtained as
\begin{align}
 \mathcal{C}_{7\gamma} &= \frac{\sqrt{2}\pi\alpha_s}{4G_F[\lambda_t]_{i3}m_{\tilde g}^2}
 \bigg[ \mathcal{R}_{ri}^{d*}\mathcal{R}_{r3}^{d} \left(\frac{8}{9}D_1(x_r)\right) 
 - \frac{m_{\tilde g}}{m_b} \mathcal{R}_{ri}^{d*}\mathcal{R}_{r6}^{d} \left(\frac{8}{9}D_2(x_r)\right) \bigg], \\
 \mathcal{C}_{8g} &= \frac{\sqrt{2}\pi\alpha_s}{4G_F[\lambda_t]_{i3}m_{\tilde g}^2}
 \bigg[ \mathcal{R}_{ri}^{d*}\mathcal{R}_{r3}^{d} \left(\frac{1}{3}D_1(x_r) - 3D_3(x_r)\right) 
 \notag \\ &\qquad\qquad\qquad\qquad
 - \frac{m_{\tilde g}}{m_b} \mathcal{R}_{ri}^{d*}\mathcal{R}_{r6}^{d} 
 \left(\frac{1}{3}D_2(x_r) - 3D_4(x_r)\right) \bigg],
 \label{eq:gluinochromo}
\end{align}
where $x_r = m_{\tilde d_r}^2/m_{\tilde g}^2$, and the loop functions are defined to be
\begin{align}
 D_1(x) &= \frac{-x^3+6x^2-3x-2-6x\ln x}{6(1-x)^4}, \\
 D_2(x) &= \frac{x^2-1-2x\ln x}{(1-x)^3}, \\
 D_3(x) &= \frac{2x^3+3x^2-6x+1-6x^2\ln x}{6(1-x)^4}, \\
 D_4(x) &= \frac{3x^2-4x+1-2x^2\ln x}{(1-x)^3}.
\end{align}
Also, $\mathcal{C}^{\prime}_{7\gamma}$ and $ \mathcal{C}^{\prime}_{8g}$ are obtained by flipping the chirality of $\mathcal{R}_{ri}^{d (*)}$ in $\mathcal{C}_{7\gamma}$ and $ \mathcal{C}_{8g}$, respectively.

In the analysis, an approximation formula in Ref.~\cite{Malm:2015oda} is used to estimate the SUSY contributions to the branching ratio of $b\to s\gamma$, where the Wilson coefficients are set at $\mu_b = 4.8$\GeV.
For $\mathcal{B}(\bar B \to X_d \gamma)$, the formula in Refs.~\cite{Hurth:2003dk,Evans:2016lzo} is used, where the SUSY contributions to the Wilson coefficients at the top-mass scale are needed. 
The latest results of the SM values are~\cite{Misiak:2015xwa}
\begin{align}
 \mathcal{B}(\bar B \to X_s \gamma)^{\rm SM} &= (3.36 \pm 0.23) \times 10^{-4}, \\
 \mathcal{B}(\bar B \to X_d \gamma)^{\rm SM} &= (1.73^{+0.12}_{-0.22}) \times 10^{-5},
\end{align}
for $E_\gamma>1.6\,{\rm GeV}$.
On the other hand, the experimental results are~\cite{Amhis:2016xyh,delAmoSanchez:2010ae,Crivellin:2011ba}
\begin{align}
 \mathcal{B}(\bar B \to X_s \gamma)^{\rm exp} &= (3.32 \pm 0.15) \times 10^{-4}, \\
 \mathcal{B}(\bar B \to X_d \gamma)^{\rm exp} &= (1.41 \pm 0.57) \times 10^{-5},
\end{align}
for $E_\gamma>1.6\,{\rm GeV}$.
In the analysis, the theoretical prediction including the SM and SUSY contributions is required to be consistent with the experimental result at the $2\sigma$ level.

$CP$ violations of $b\to d_i\gamma$ are sensitive to the imaginary parts of flavor-violating scalar trilinear couplings.
Long-distance effects tend to spoil the sensitivity~\cite{Benzke:2010tq}.
This could be resolved by taking a difference of the \CP asymmetries~\cite{Benzke:2010tq},
\begin{align}
 \Delta A_{\rm CP}(b\to s\gamma) &= 
 A_{\rm CP}(B^-\to X_s^- \gamma)-A_{\rm CP}(\bar B^0\to X_s^0 \gamma) 
 \notag \\ 
 &= 4\pi^2\alpha_s(\mu_b)\,\frac{\widetilde\Lambda_{78}}{m_b}\,{\rm Im}
 \left[ \frac{\mathcal{C}_{7\gamma}^* \mathcal{C}_{8g}
  + \mathcal{C}_{7\gamma}^{\prime*} \mathcal{C}'_{8g}}{|\mathcal{C}_{7\gamma}|^2+|\mathcal{C}'_{7\gamma}|^2} \right],
 \label{eq:DACP}
\end{align}
where the right-handed contributions are taken into account~\cite{Kagan:1998bh}.
The hadronic parameter $\widetilde\Lambda_{78}$ introduces an uncertainty to the analysis and is estimated to be $12\MeV < \widetilde\Lambda_{78} < 190\MeV$~\cite{Malm:2015oda}. 
We take an average value, $\widetilde\Lambda_{78} = 89\MeV$, in the analysis.
The Wilson coefficients include both the SM and SUSY contributions, which are evaluated at the scale $\mu_b = 2\GeV$.
The SM prediction is expected to be much suppressed, $\Delta A_{\rm CP}(b\to s\gamma)^{\rm SM} \approx 0$
~\cite{Benzke:2010tq}.
On the other hand, the experimental result is~\cite{Lees:2014uoa}
\begin{align}
 \Delta A_{\rm CP}(b\to s\gamma)^{\rm exp} = (5.0 \pm 3.9_{\rm stat} \pm 1.5_{\rm syst})\%
 \label{eq:ExpDAcp}
\end{align}
from the BaBar experiment. 
The Belle experiment also published a result on $\Delta A_{\rm CP}(B \to K^*\gamma)$~\cite{Horiguchi:2017ntw},
\begin{align}
 \Delta A_{\rm CP}(B \to K^*\gamma)^{\rm exp} = (2.4 \pm2.8_{\rm stat} \pm 0.5_{\rm syst})\%.
\end{align}
The asymmetry of the inclusive decay is expected to be comparable to that of the exclusive mode~\cite{Ishikawa}.
Both results are consistent with a null asymmetry difference. 
Since the uncertainties are large, the SUSY parameters will not be constrained in the region of our interest.
In future, the uncertainty is projected to achieve 0.37\% for $\Delta A_{\rm CP}(b\to s\gamma)$ at Belle II with $50\,{\rm ab}^{-1}$~\cite{Sandilya:2017mkb}.\footnote{
Although the experimental uncertainty of the direct \CP asymmetry $A_{\rm CP}(b \to s\gamma)$ is also projected to be sub-percent level~\cite{Sandilya:2017mkb}, 
long-distance contributions as well as hadronic uncertainties spoil the SM prediction~\cite{Benzke:2010tq}.
}

\section{Vacuum stability}
\label{sec:vacuum}

The Wilson coefficients in Eqs.~\eqref{eq:CHQ}--\eqref{eq:CHD} are enhanced by large off-diagonal trilinear couplings, $\left(T_D\right)_{i3}$ and $\left( T_D\right)_{3i}$ $(i=1,2)$.
Such large trilinear couplings tend to generate dangerous charge and color breaking (CCB) global minima in the scalar potential \cite{Park:2010wf}. 
Hence, they are limited by the vacuum (meta-)stability condition: the lifetime of the EW vacuum must be longer than the age of the Universe.
In this section, we will investigate the vacuum stability conditions of $\left(T_D\right)_{i3}$ and $\left( T_D\right)_{3i}$.

The vacuum decay rate  per unit volume is represented by $\Gamma/V = A \exp \left( - S_E \right)$, where $S_E$ is the Euclidean action of the bounce solution~\cite{Coleman:1977py}.
\texttt{CosmoTransition} 2.0.2~\cite{Wainwright:2011kj} is used to estimate $S_E$ at the semiclassical level.
The prefactor $A$ cannot be determined unless radiative corrections are taken into account~\cite{Callan:1977pt, Endo:2015ixx}. 
We adopt an order-of-magnitude estimation, $A \sim \left( 100\GeV\right)^4$.
By requiring $(\Gamma/V)^{1/4}$ to be smaller than the current Hubble parameter, the lifetime of the EW vacuum becomes longer than the age of the Universe. The condition corresponds to $S_E \gtrsim 400$.
In this paper, thermal effects and radiative corrections to the vacuum transitions are discarded.

The bounce solution and $S_E$ are determined by the scalar potential.
The potential relevant for the vacuum decay generated by $\left(T_D\right)_{13}$ and/or $\left(T_D\right)_{31}$ is
\begin{align}
 V &=  
 \frac{1}{2} m_{11}^2 \, h_d^2 
 + \frac{1}{2} m_{22}^2 \, h_u^2 
 - m_{12}^2 \, h_d h_u 
 \notag \\ &~~
 + \frac{1}{2} m_{\tilde Q,1}^2 \,\tilde d_L^2 
 + \frac{1}{2} m_{\tilde Q,3}^2 \,\tilde b_L^2 
 + \frac{1}{2} m_{\tilde D,1}^2 \,\tilde d_R^2
 + \frac{1}{2} m_{\tilde D,3}^2 \,\tilde b_R^2 
 \notag \\ &~~ 
 + \frac{1}{\sqrt{2}} \left[ \left(T_D\right)_{33} h_d - y_b \mu h_u \right]\tilde b_L \tilde b_R 
 + \frac{1}{\sqrt{2}} \left(T_D \right)_{13}  h_d \tilde d_L \tilde b_R 
 + \frac{1}{\sqrt{2}} \left(T_D \right)_{31} h_d \tilde b_L \tilde d_R 
 \notag \\ & ~~
 + \frac{1}{4} y_b^2 (\tilde b_L^2 \tilde b_R^2 + \tilde b_L^2 h_d^2 + \tilde b_R^2 h_d^2) 
 \notag \\ &~~
 + \frac{1}{24} g_3^2 (\tilde d_L^2 + \tilde b_L^2 - \tilde d_R^2 - \tilde b_R^2)^2
 + \frac{1}{32} g_2^2 (h_u^2 - h_d^2 + \tilde d_L^2 + \tilde b_L^2)^2
 \notag \\ &~~
 + \frac{1}{32} g_Y^2 \left(h_u^2 - h_d^2 + \frac{1}{3}\tilde d_L^2 + \frac{1}{3}\tilde b_L^2 + \frac{2}{3}\tilde d_R^2  + \frac{2}{3}\tilde b_R^2 \right)^2,
 \label{eq:scalarPT}
\end{align}
where the coefficients are 
\begin{align}
 m_{11}^2 &= m_A^2 \sin^2\beta - \frac{1}{2} m_Z^2 \cos 2\beta, \\
 m_{22}^2 &= m_A^2 \cos^2\beta + \frac{1}{2} m_Z^2 \cos 2\beta, \\
 m_{12}^2 &= \frac{1}{2} m_A^2 \sin 2\beta.
\end{align}
Here, $h_d$, $h_u$, $\tilde d_L$, $\tilde b_L$, $\tilde d_R$, $\tilde b_R$ are real scalar fields with $\langle h_d \rangle  = v \cos\beta $ and $\langle h_u \rangle  = v \sin\beta $ at the EW vacuum. 
In this potential, all coefficients can be rotated to be real by rephasing the fields.
The terms proportional to light flavor Yukawas are discarded, because those contributions are negligible.  
The scalar potential for $\tilde s_{L}$, $\tilde s_{R}$ is obtained by substituting $\tilde d_{L,R}$ $\to $ $\tilde s_{L,R}$, $\left(T_D\right)_{13}$ $\to $ $\left(T_D\right)_{23}$, and $\left(T_D\right)_{31}$ $\to $ $\left(T_D\right)_{32}$.

\begin{figure}[t]
\begin{center}
\includegraphics[width=0.6\textwidth, bb= 0 0 546 343]{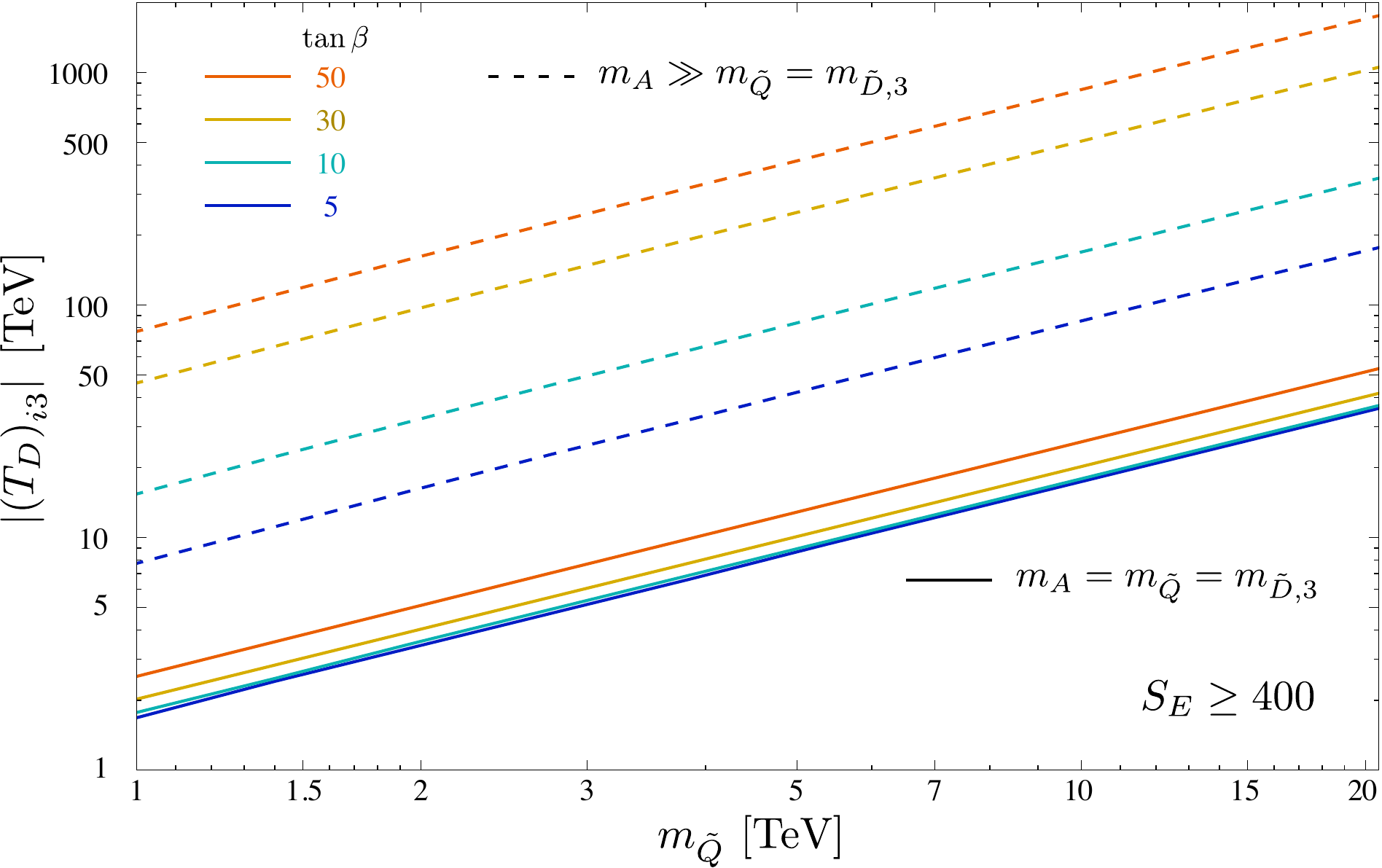}
\caption{
The upper bound on $\left| \left(T_D \right)_{i3} \right|$ for $i=1,2$ from the vacuum stability condition as a function of $m_{\tilde{Q}}  $. Here, $\tan \beta =5,$ 10, 30, 50 are taken.
The solid lines are in the case of $m_A = m_{\tilde{Q},i}  = m_{\tilde{D},3} \equiv m_{\tilde{Q}}$, while the dashed lines  represent the decoupling limit of the heavy Higgs multiplets, $ m_A \gg m_{\tilde{Q},i}  = m_{\tilde{D},3} \equiv m_{\tilde{Q}}$.
}
\label{fig:TD13bound}
\end{center}
\end{figure}

Let us first consider the vacuum stability condition when only $\left(T_D\right)_{13}$ is large.
The scalar potential is simplified to be
\begin{align}
 V &= 
 \frac{1}{2} m_{11}^2 \, h_d^2 
 + \frac{1}{2} m_{22}^2 \, h_u^2 
 - m_{12}^2 \, h_d h_u 
 + \frac{1}{2} m_{\tilde Q,1}^2 \,\tilde d_L^2 
 + \frac{1}{2} m_{\tilde D,3}^2 \,\tilde b_R^2 
 + \frac{1}{\sqrt{2}} \left(T_D\right)_{13} h_d \tilde d_L \tilde b_R 
 \\ &~~ 
 + \frac{1}{4} y_b^2 \tilde b_R^2 h_d^2
 + \frac{1}{24} g_3^2 (\tilde d_L^2 - \tilde b_R^2)^2
 + \frac{1}{32} g_2^2 (h_u^2 - h_d^2 + \tilde d_L^2)^2
 + \frac{1}{32} g_Y^2 \left(h_u^2 - h_d^2 + \frac{1}{3}\tilde d_L^2 + \frac{2}{3}\tilde b_R^2 \right)^2.
 \notag
\end{align}
When $m_A \sim m_{\tilde{Q},1} \sim m_{\tilde{D},3}$, CCB vacua appear around a $h_d$--$\tilde d_L$--$\tilde b_R$ plane.
In Fig.~\ref{fig:TD13bound}, the solid lines show upper bounds on $\left| \left(T_D \right)_{13} \right|$ for $\tan \beta = 5$, 10, 30, and 50. 
We assumed $m_A = m_{\tilde{Q},1} = m_{\tilde{D},3}$. 
It is shown that the upper bounds are proportional to $m_{\tilde Q}$.
Also, the results depend on $\tan \beta$ slightly.
This is because the scalar potential is stabilized by a quartic coupling $y_b^2 \tilde b_R^2 h_d^2 \sim  \left( 2 m_b^2 /v^2\right) \tan^2 \beta  \tilde b_R^2 h_d^2 $, when $\tan \beta$ is large.

\begin{figure}[t]
\begin{center}
\includegraphics[width=0.6\textwidth, bb= 0 0 341 216]{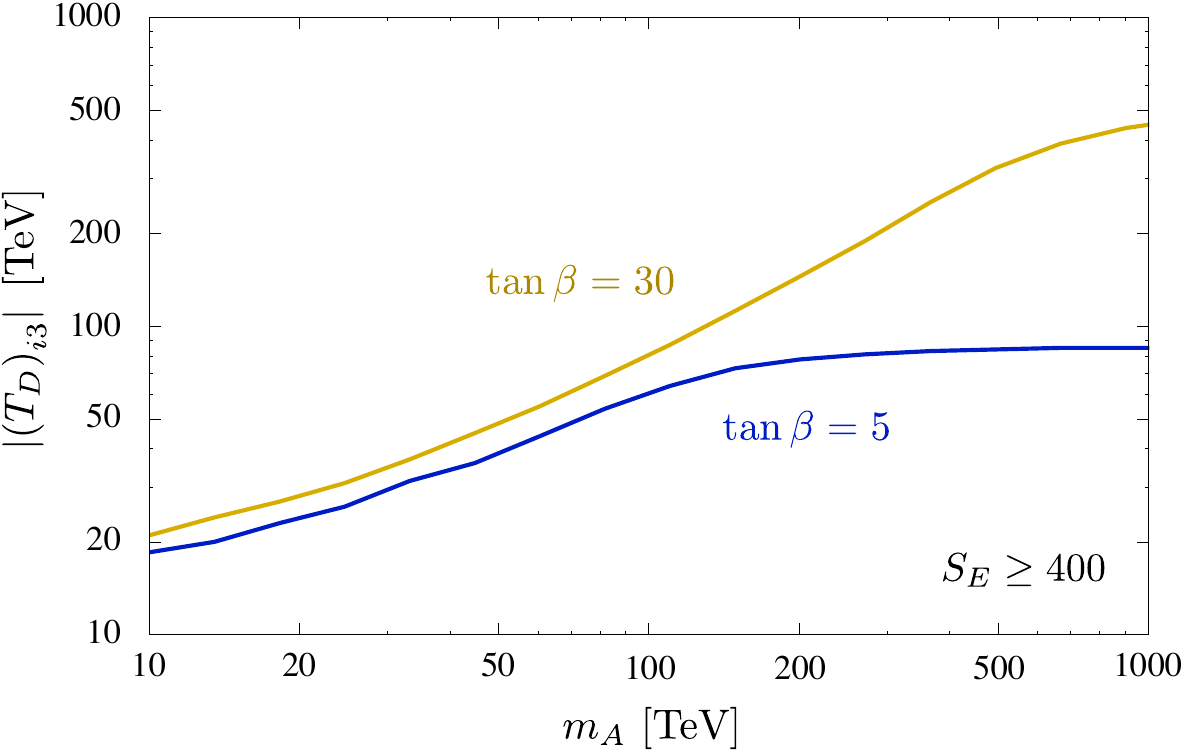}
\caption{The vacuum stability condition of  $\left| \left(T_D \right)_{i3} \right|$ for $i=1,2$ as a function of $m_A$. Here,  $m_{\tilde{Q},i}  = m_{\tilde{D},3} = 10\TeV$, and 
$\tan \beta$ = 5 and 30 are taken.
}
\label{fig:TD13_MA}
 \end{center}
\end{figure}

When $m_A$ is larger than $m_{\tilde{Q},1} \sim m_{\tilde{D},3}$, the position of the CCB vacuum approaches to a $H$--$\tilde d_L$--$\tilde b_R$ plane, where $H$ includes the SM-like Higgs boson, $H = h_{\rm SM}+v$. 
In Fig.~\ref{fig:TD13_MA}, the $m_A$ dependence of the upper bound is shown.
Here, $\tan \beta =5$ and 30 are taken.
We found that the vacuum stability condition is relaxed for large $m_A$.

In the decoupling limit of the heavy Higgs bosons ($m_A^2 \gg m_Z^2, \alpha \to \beta - \pi/2$), the scalar potential can be expressed by $H$, $\tilde d_L$, and $\tilde b_R$ as 
\begin{align}
 V &= 
 - \frac{1}{4} m_Z^2 \cos^2 2\beta\, H^2 
 + \frac{1}{2} m_{\tilde Q,1}^2 \,\tilde d_L^2 
 + \frac{1}{2} m_{\tilde D,3}^2 \,\tilde b_R^2 
 + \frac{1}{\sqrt{2}} \left(T_D\right)_{13} \cos\beta\, H \tilde d_L \tilde b_R 
 \notag \\ &~~
 + \frac{1}{4} y_b^2 \tilde b_R^2 H^2 \cos^2\beta 
 + \frac{1}{24} g_3^2 (\tilde d_L^2 - \tilde b_R^2)^2
 + \frac{1}{32} g_2^2 (H^2 \cos2\beta - \tilde d_L^2)^2
 \notag \\ &~~
 + \frac{1}{32} g_Y^2 \left( H^2 \cos2\beta - \frac{1}{3}\tilde d_L^2 - \frac{2}{3}\tilde b_R^2 \right)^2.
 \label{eq:Vdecoupled}
\end{align}
The upper bounds on $\left| \left(T_D \right)_{13} \right|$ are shown by the dashed lines in Fig.~\ref{fig:TD13bound}.\footnote{
In this scalar potential, the SM-like Higgs boson is lighter than 125\GeV.
The vacuum stability condition can be evaluated naively by adding top-stop radiative corrections, 
$\left(g_2^2 + g_Y^2 \right) \delta_H^{(t)} \sin^4 \beta H^4 /8 $,~\cite{Hisano:2010re, Kitahara:2012pb, Kitahara:2013lfa, Carena:2012mw} to Eq.~\eqref{eq:Vdecoupled} 
in order to achieve the 125\GeV~SM-like Higgs boson at the EW vacuum.
We found that Eq.~\eqref{eq:vacuumfit} is barely changed.
Dedicated studies are needed to fully include the radiative corrections (see Ref.~\cite{Endo:2015ixx}).
}
Again, they are proportional to $m_{\tilde Q}$.
In contrast to the case of $m_A \sim m_{\tilde{Q}}$, the result is almost proportional to $\tan \beta$.
This is understood by $\cos \beta$ associated to $\left(T_D\right)_{13}$. 
A fitting formula of the vacuum stability condition in the large $m_A$ limit with $m_{\tilde{Q},1} = m_{\tilde{D},3} \equiv m_{\tilde{Q}}$ is derived as 
\beq
\frac{\left| \left(T_D\right)_{13}\right| }{\tan \beta} \lesssim - 0.186 \TeV+ 1.675\, m_{\tilde{Q}},
\label{eq:vacuumfit}
\eeq
where the phase of $\left(T_D\right)_{i3}$ is taken into account.
This formula works well for $m_{\tilde{Q}} > 1\TeV$. 

Let us next turn on $\left(T_D\right)_{23}$ in addition to $\left(T_D\right)_{13}$.
The scalar trilinear term becomes
\begin{align}
 V \supset \frac{1}{\sqrt{2}} 
 \left[ \left(T_D\right)_{13} \tilde d_L + \left(T_D\right)_{23} \tilde s_L \right] \tilde b_R h_d.
\end{align}
Here, $\left(T_D\right)_{13,23}$ are taken to be real by rephasing the scalar fields. 
By mixing $\tilde d_L$ and $\tilde s_L$, one can obtain
\begin{align}
 V \supset \frac{1}{\sqrt{2}} 
 \left[\left(T_D\right)_{13}^2 + \left(T_D\right)_{23}^2\right]^{1/2}\, \tilde d_L' \tilde b_R h_d,
\end{align}
where $\tilde d_L = \tilde d_L' \cos\theta - \tilde s_L'\sin\theta$ and $\tilde s_L = \tilde d_L' \sin\theta + \tilde s_L'\cos\theta$ with $\tan\theta = \left(T_D\right)_{23}/\left(T_D\right)_{13}$.
When $m_{\tilde Q,1}^2 = m_{\tilde Q,2}^2 \equiv m_{\tilde Q}^2$, the scalar potential of $\tilde d_L'$ is obtained from that of $\tilde d_L$ by substituting $\left(T_D\right)_{13} \to \left[\left(T_D\right)_{13}^2 + \left(T_D\right)_{23}^2\right]^{1/2}$ as well as $\tilde d_L \to \tilde d_L'$.
Therefore, the vacuum stability condition \eqref{eq:vacuumfit} is extended to be
\beq
\frac{ \sqrt{|\left(T_D\right)_{13}|^2 + |\left(T_D\right)_{23}|^2} }{\tan \beta} \lesssim - 0.186 \TeV+ 1.675\, m_{\tilde{Q}},
\label{eq:vacuumfit2}
\eeq
where the phases of $\left(T_D\right)_{13,23}$ are taken into account appropriately.
The formula is valid when $m_{\tilde{Q}} \equiv m_{\tilde{Q},1} = m_{\tilde{Q},2} = m_{\tilde{D},3} > 1\TeV$ and $m_A$ is decoupled.\footnote{
 We have validated the formula \eqref{eq:vacuumfit2} explicitly by analyzing the bounce action of the scalar potential of $H$, $\tilde d_L$, $\tilde s_L$, and $\tilde b_R$. }

When only $\left(T_D\right)_{31}$ is large, the potential becomes
\begin{align}
 \label{eq:potential2}
 V &= 
 \frac{1}{2} m_{11}^2 \, h_d^2 
 + \frac{1}{2} m_{22}^2 \, h_u^2 
 - m_{12}^2 \, h_d h_u 
 + \frac{1}{2} m_{\tilde Q,3}^2 \,\tilde b_L^2 
 + \frac{1}{2} m_{\tilde D,1}^2 \,\tilde d_R^2
 + \frac{1}{\sqrt{2}} \left(T_D\right)_{31} h_d \tilde b_L \tilde d_R 
 \\ &~~
 + \frac{1}{4} y_b^2 \tilde b_L^2 h_d^2
 + \frac{1}{24} g_3^2 (\tilde b_L^2 - \tilde d_R^2)^2
 + \frac{1}{32} g_2^2 (h_u^2 - h_d^2 + \tilde b_L^2)^2
 + \frac{1}{32} g_Y^2 \left( h_u^2 - h_d^2 + \frac{1}{3}\tilde b_L^2 + \frac{2}{3}\tilde d_R^2 \right)^2.
 \notag
\end{align}
By repeating the above procedure, one can obtain quantitatively the same fitting formula for $\left(T_D\right)_{3i}$ as Eq.~\eqref{eq:vacuumfit2}, 
\beq
\frac{ \sqrt{|\left(T_D\right)_{31}|^2 + |\left(T_D\right)_{32}|^2} }{\tan \beta} \lesssim - 0.186 \TeV+ 1.675\, m_{\tilde{Q}},
\label{eq:vacuumfit3}
\eeq
where $m_{\tilde{Q}} \equiv m_{\tilde{Q},3} = m_{\tilde{D},1} =  m_{\tilde{D},2} > 1\TeV$ and $m_A$ is decoupled.

\section{Numerical analysis}
\label{sec:analysis}

In this section, we study gluino contributions to \epoe\,via the $Z$ penguin. 
They are enhanced by large scalar trilinear couplings as shown in Sec.~\ref{sec:SUSY}.
Since $(T_D)_{13,23,31,32}$ are complex variables, there are 8 degrees of freedom. 
For simplicity, we restrict the parameter space such that two of $(T_D)_{13,23,31,32}$ are real.
When $(T_D)_{23,32}$ are real, we checked that wide parameter regions to explain the discrepancy of \epoe\,are tightly excluded by $\mathcal{B}(\bar B \to X_{d,s} \gamma)$.
Therefore, we consider the cases when $(T_D)_{13,31}$ are real. 
The scalar trilinear coupling are parameterized as
\begin{align}
 [(T_D)_{13},(T_D)_{23},(T_D)_{31},(T_D)_{32}] = 
 [\gamma_L,\alpha_L + i \beta_L,\gamma_R,\alpha_R + i \beta_R],
\label{eq:TDset}
\end{align}
where $\alpha_i$, $\beta_i$ and $\gamma_i$ are real parameters. 
Then, one obtains (see Sec.~\ref{sec:SUSY})
\begin{align}
  {\rm Im}\,[\mathcal{C}_{HQ}^{(1,3)}]_{12} &\propto 
  -{\rm Im}\,\left[(T_D)_{13}^* (T_D)_{23}\right] = -\beta_L \gamma_L, \\
  {\rm Im}\,[\mathcal{C}_{HD}]_{12} &\propto 
  +{\rm Im}\,\left[(T_D)_{31} (T_D)_{32}^* \right]= -\beta_R \gamma_R.
\end{align}
The $L$ variables contribute to the left-handed Wilson coefficients, and the $R$ variables to the right-handed ones.
In order to evaluate the observables, we scan the whole parameter region of $\alpha_i$, $\beta_i$, and $\gamma_i$ where the vacuum stability conditions are satisfied.\footnote{
 We checked that the constraint from $\mathcal{B}(K_L\to\mu^+\mu^-)$ is weaker than the other constraints in the parameter region of our interest.
}

When $\beta_L\gamma_L > 0$ and $\beta_R\gamma_R > 0$, the SUSY contribution to \epoe\,is maximized, because the left-handed contribution, $\mathcal{C}_{HQ}$, constructively interferes with the right-handed one, $\mathcal{C}_{HD}$.
In this case, $\mathcal{B}(K_L\to\pi^0\nu\bar\nu)$ cannot exceed the SM prediction, because positive $\beta_L\gamma_L$ and $\beta_R\gamma_R$ tends to decrease the branching ratio, as can be seen from Eq.~\eqref{eq:KLpinn}.
We consider this case in Sec.~\ref{Sec:posiposi}.
In contrast, \epoe\, cannot be accommodated with the result \eqref{eq:eps_limit} for $\beta_L\gamma_L < 0$ and $\beta_R\gamma_R < 0$. When either $\beta_L\gamma_L$ or $\beta_R\gamma_R$ is negative, the discrepancy of \epoe\, can also be explained. Because the right-handed contribution to \epoe\,is larger than the left-handed one, $\beta_R\gamma_R > 0$ is favored to amplify \epoe. At the same time, $\mathcal{B}(K_L\to\pi^0\nu\bar\nu)$ can be enhanced and may exceed the SM value. 
Hence, we consider the case when $\beta_L\gamma_L < 0$ and $\beta_R\gamma_R > 0$ in Sec.~\ref{Sec:negaposi}.

Before proceeding to the analysis, let us summarize assumptions on model parameters.
Since the vacuum stability condition is relaxed by large $m_A$, the heavy Higgs bosons are supposed to be decoupled. 
The squark masses are set to be degenerate, $m_{\tilde{Q}} \equiv m_{\tilde{Q},1} = m_{\tilde{Q},2} = m_{\tilde{Q},3} = m_{\tilde{D},1} = m_{\tilde{D},2} =  m_{\tilde{D},3}$, for simplicity.
The Higgsino mass parameter is also equal to $m_{\tilde Q}$, though dependences of the observables on it are weak.
We take $\tan\beta=5$, though the following results are insensitive to the choice, because the observables as well as the vacuum stability condition depend on it dominantly in a combination of $T_D\cos\beta$.

\subsection{$\boldsymbol{\beta_L\gamma_L > 0}$ and $\boldsymbol{\beta_R\gamma_R > 0}$}
\label{Sec:posiposi}

\begin{figure}[t!]
\begin{center}
\includegraphics[scale=0.5, bb= 0 0 360 363]{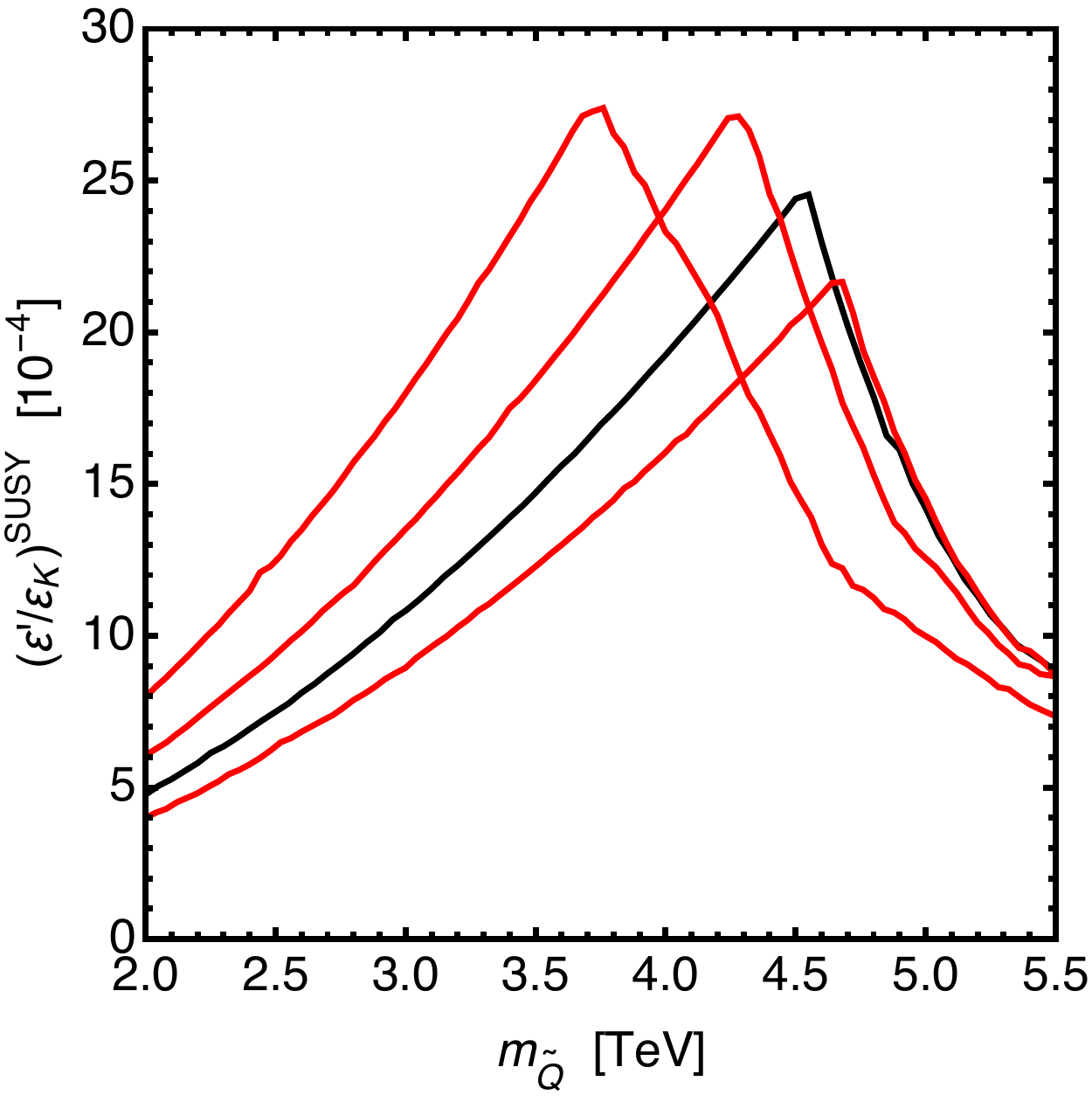}\hspace{4mm}
\includegraphics[scale=0.5, bb= 0 0 360 369]{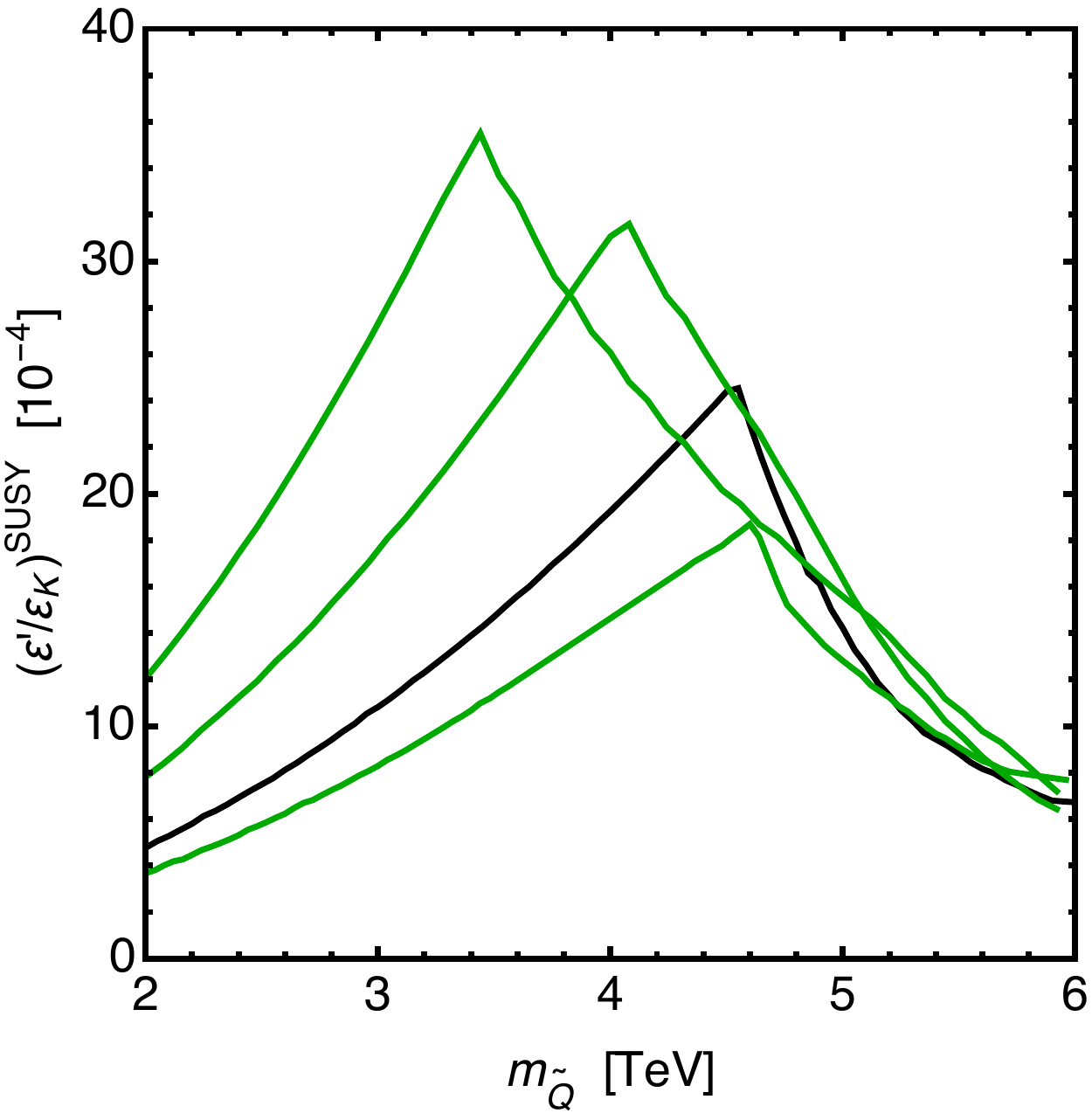}
\caption{
The maximal gluino contributions to \epoe\,as a function of $m_{\tilde{Q}}$. The parameters are $\gamma_R/\beta_R=\gamma_L/\beta_L=1$ and $m_{\tilde{g}}/m_{\tilde{Q}}=1$ on the black line. In the left plot, $\gamma_R/\beta_R=\gamma_L/\beta_L=0.6, 0.8, 1.2$ with $m_{\tilde{g}}/m_{\tilde{Q}}=1$ from left to right of the red lines. In the right plot, $m_{\tilde{g}}/m_{\tilde{Q}}=1.8, 1.4, 0.8$ with $\gamma_R/\beta_R=\gamma_L/\beta_L=1$ from left to right of the green lines. 
}
\label{fig:gluinog12}
\end{center}
\end{figure}

In Fig.~\ref{fig:gluinog12}, the maximal values of the SUSY contributions to \epoe\,are shown for $\beta_L\gamma_L > 0$ and $\beta_R\gamma_R > 0$ as a function of $m_{\tilde{Q}}$.
There is a peak structure for each line.
In smaller squark mass regions, the maximal value is determined by $\mathcal{B}(\bar B \to X_d \gamma)$.
Defining the squark mixing parameter, $\delta_D = (T_D)_{ij}v\cos\beta /m_{\tilde Q}^2$, the SUSY contributions to \epoe\,depend on it as $( \epoe )^{\rm SUSY} \sim \delta_D^2$, whereas those to $\mathcal{B}(\bar B \to X_d \gamma)$ is $\sim \delta_D/m_{\tilde Q}$, where $m_{\tilde g} \sim m_{\tilde Q}$ is supposed.
Thus, the maximal value of \epoe\,increases as $m_{\tilde Q}$ becomes larger.
In larger squark mass regions, the maximal value is determined by \epsk, $\mathcal{B}(\bar B \to X_s \gamma)$ and the vacuum stability condition as well as $\mathcal{B}(\bar B \to X_d \gamma)$.
In particular, the gluino box contribution to \epsk\,depends on $\delta_D$ as $\sim \delta_D^4/m_{\tilde Q}^2$, whereas the SUSY contributions via $\mathcal{C}_{HQ}$ and $\mathcal{C}_{HD}$ are not suppressed by $m_{\tilde Q}$, i.e., behaves as $\sim \lambda_t \delta_D^2 / m_Z^2$.
When $m_{\tilde Q}$ is small, the latter contribution can be canceled enough by the former one.
However, as $m_{\tilde Q}$ increases, the cancellation becomes weaker in the parameter region allowed by the other constraints.
Hence, the bounds on the trilinear couplings become severer to satisfy the constraint of \epsk. 
Consequently, the maximal value of \epoe\,decreases. 

In the figures, $\gamma_i/\beta_i$ or $m_{\tilde{g}}/m_{\tilde{Q}}$ is also varied. 
On the black line, $\gamma_R/\beta_R=\gamma_L/\beta_L=1$ and $m_{\tilde{g}}/m_{\tilde{Q}}=1$ are chosen. 
In the left plot, $\gamma_R/\beta_R=\gamma_L/\beta_L=0.6, 0.8, 1.2$ with $m_{\tilde{g}}/m_{\tilde{Q}}=1$ from left to right of the red lines. 
On the other hand, $m_{\tilde{g}}/m_{\tilde{Q}}=1.8, 1.4, 0.8$ with $\gamma_R/\beta_R=\gamma_L/\beta_L=1$ from left to right of the green lines in the right plot. 
The maximum value increases when $\gamma_i/\beta_i$ is small and $m_{\tilde{g}}/m_{\tilde{Q}}$ is large. 
Also, it is found that the current discrepancy of \epoe\,can be explained if the squark mass is smaller than $5.6~\mbox{TeV}$.

\subsection{$\boldsymbol{\beta_L\gamma_L < 0}$ and $\boldsymbol{\beta_R\gamma_R > 0}$}
\label{Sec:negaposi}

\begin{figure}[t!]
\begin{center}
\includegraphics[scale=0.5, bb= 0 0 360 351]{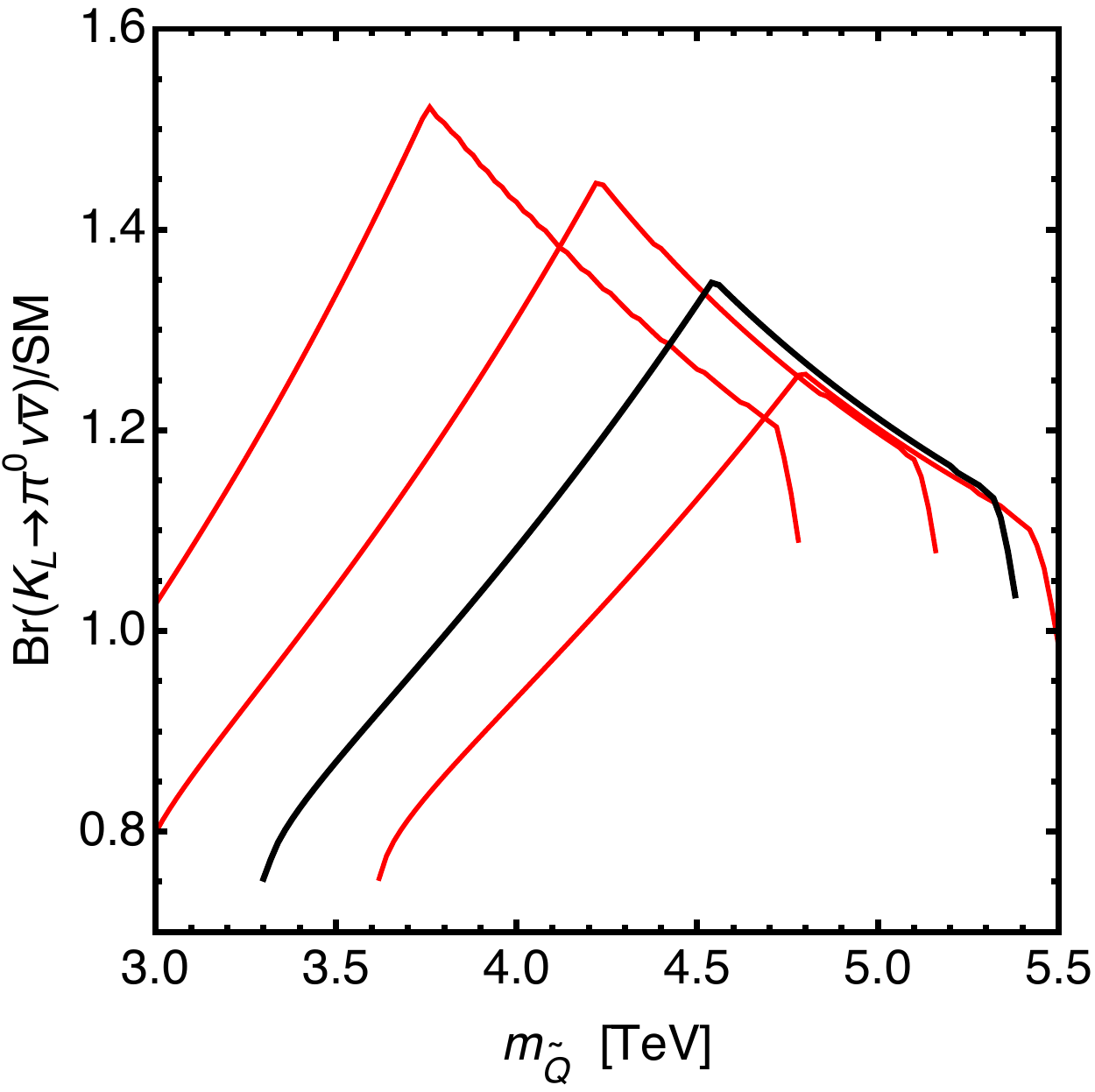}\hspace{4mm}
\includegraphics[scale=0.5, bb= 0 0 360 369]{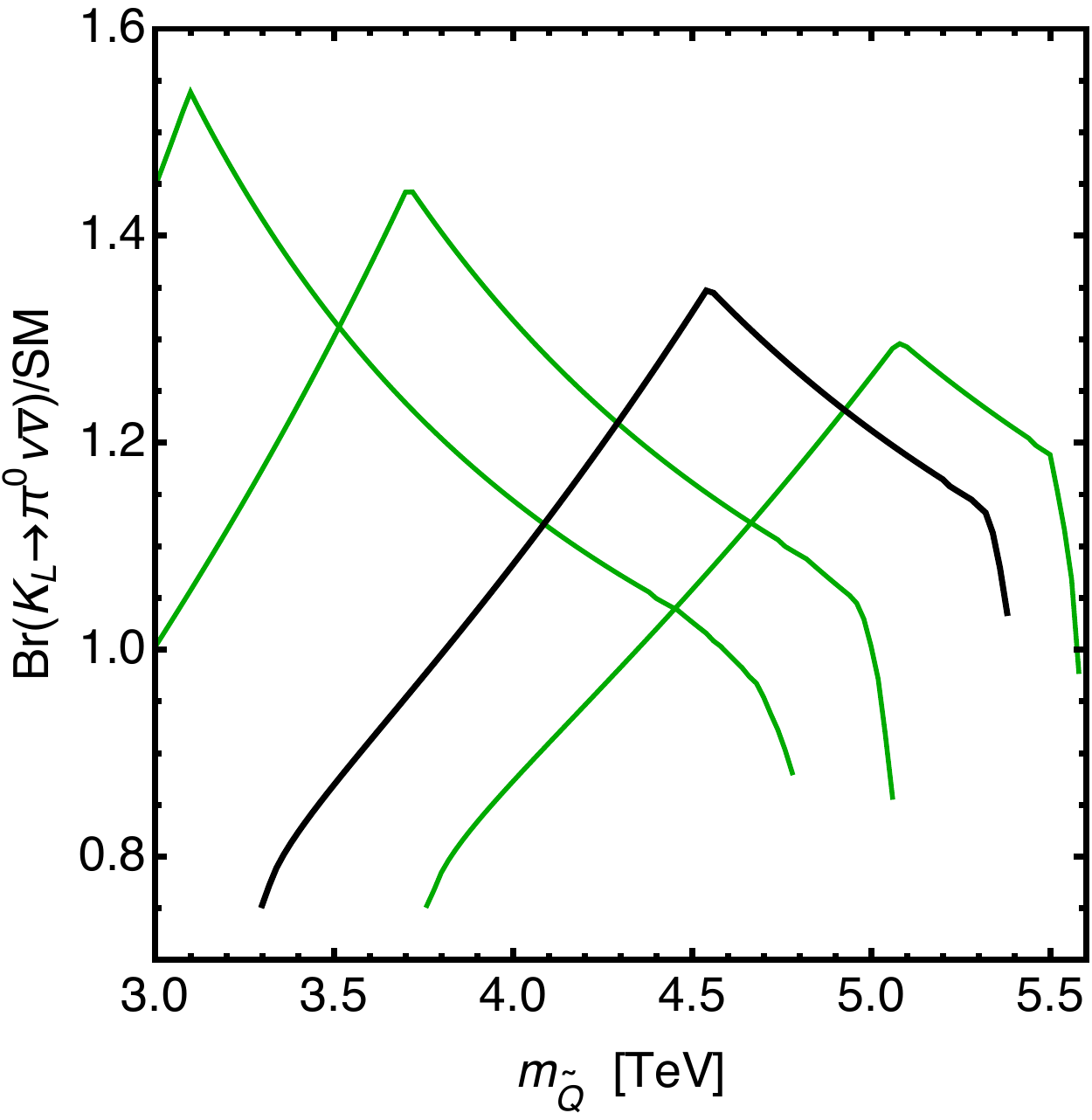}
\caption{
The maximum value of $\mathcal{B}(K_L\to\pi^0\nu\bar\nu)$ normalized by the SM prediction as a function of $m_{\tilde Q}$.
Here, $( \epoe )^{\rm SUSY} = 10.0 \times 10^{-4}$ is fixed.
The parameters are $\gamma_R/\beta_R = -\gamma_L/\beta_L = 1$ and $m_{\tilde g}/m_{\tilde Q} = 1$ on the black line. 
In the left plot, $\gamma_R/\beta_R = -\gamma_L/\beta_L = 0.6, 0.8, 1.2$ with $m_{\tilde g}/m_{\tilde Q} = 1$ from left to right of the red lines.
In the right plot, $m_{\tilde g}/m_{\tilde Q} = 1.8, 1.4, 0.8$ with $\gamma_R/\beta_R = -\gamma_L/\beta_L = 1$ from left to right of the green lines.
}
\label{fig:KLpinn}
\end{center}
\end{figure}

We study other observables with keeping the SUSY contribution to \epoe\,sizable for $\beta_L\gamma_L < 0$ and $\beta_R\gamma_R > 0$.
The SUSY parameters are determined to achieve $( \epoe )^{\rm SUSY} = 10.0 \times 10^{-4}$, where  the current discrepancy between the experimental and SM values is explained at the $1\sigma$ level. 

In Fig.~\ref{fig:KLpinn}, $\mathcal{B}(K_L\to\pi^0\nu\bar\nu)$ is maximized for given $m_{\tilde Q}$.
One finds a peak structure for each line.
On the left side of the peak, the parameters are constrained by $\mathcal{B}(\bar B \to X_d \gamma)$.
If the soft masses are too small, \epoe\,cannot be large sufficiently. 
On the right side, the constraints from \epsk\,and $\mathcal{B}(\bar B \to X_s \gamma)$ become relevant.
When SUSY particles are very heavy, the SUSY contribution to \epsk\,via $\mathcal{C}_{HQ}$ and $\mathcal{C}_{HD}$ cannot be canceled enough by that via the gluino box contribution in the parameter region allowed by the other constraints.

One can see that $\mathcal{B}(K_L\to\pi^0\nu\bar\nu)$ can be larger than the SM value.
This result is contrasted with the case when $\beta_L\gamma_L > 0$ and $\beta_R\gamma_R > 0$.

In the figures, $\gamma_i/\beta_i$ or $m_{\tilde g}/m_{\tilde Q}$ is also varied.
On the black line, $\gamma_R/\beta_R = -\gamma_L/\beta_L = 1$ and $m_{\tilde g}/m_{\tilde Q} = 1$ are chosen.
In the left plot, $\gamma_R/\beta_R = -\gamma_L/\beta_L = 0.6, 0.8, 1.2$ with $m_{\tilde g}/m_{\tilde Q} = 1$ from left to right of the red lines.
On the other hand, $m_{\tilde g}/m_{\tilde Q} = 1.8, 1.4, 0.8$ with $\gamma_R/\beta_R = -\gamma_L/\beta_L = 1$ from left to right of the green lines in the right plot.
In both plots, the peak positions depend on the setup.
The maximum value increases when $|\gamma_i/\beta_i|$ is small and/or $m_{\tilde g}/m_{\tilde Q}$ is large.
It is found that $\mathcal{B}(K_L\to\pi^0\nu\bar\nu)$ can be about 1.5 times larger than the SM prediction.
Such a branching ratio could be discovered in future KOTO experiment.

\begin{figure}[t]
\begin{center}
\includegraphics[scale=0.5, bb= 0 0 360 371]{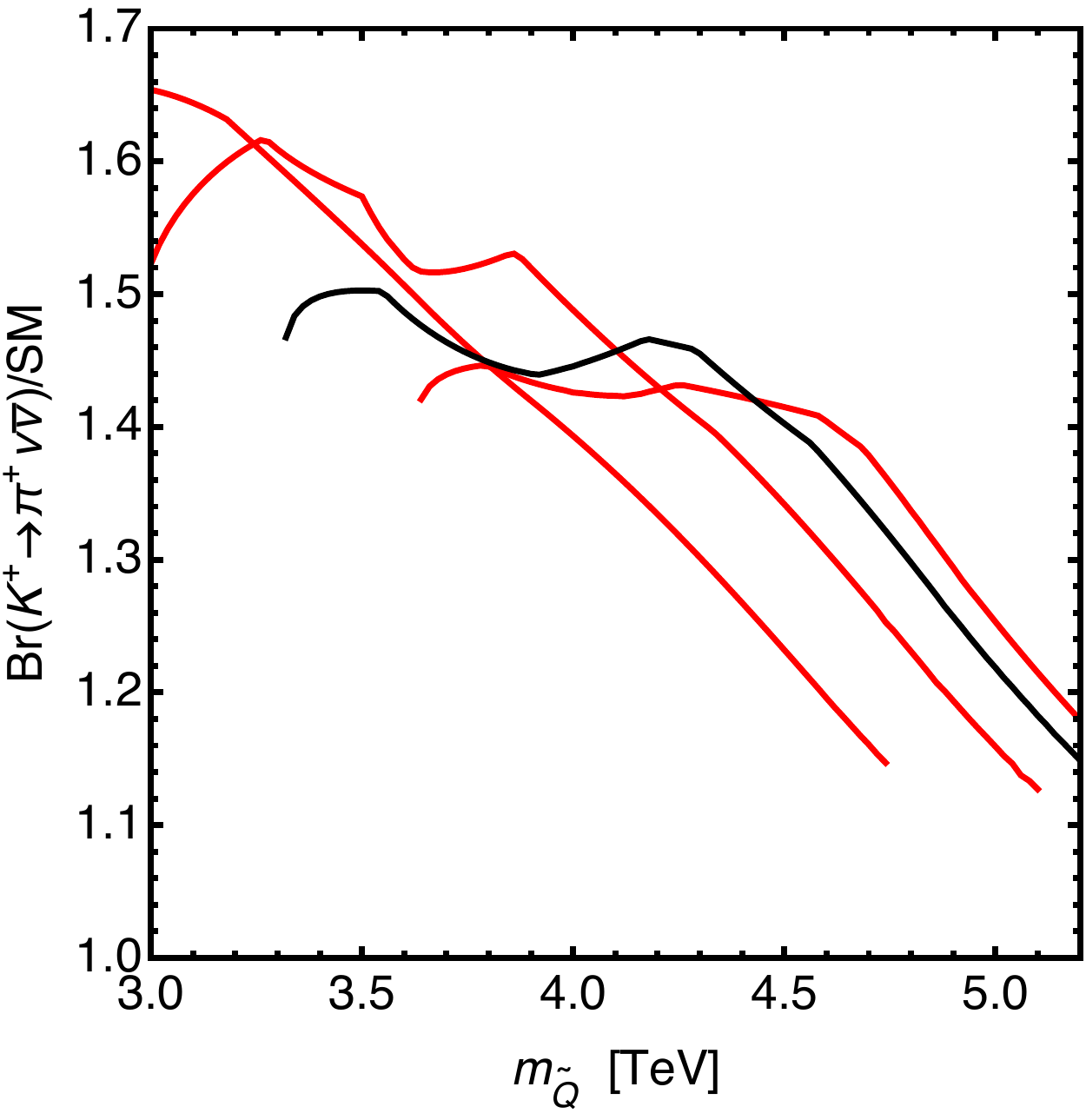}\hspace{4mm}
\includegraphics[scale=0.5, bb= 0 0 360 370]{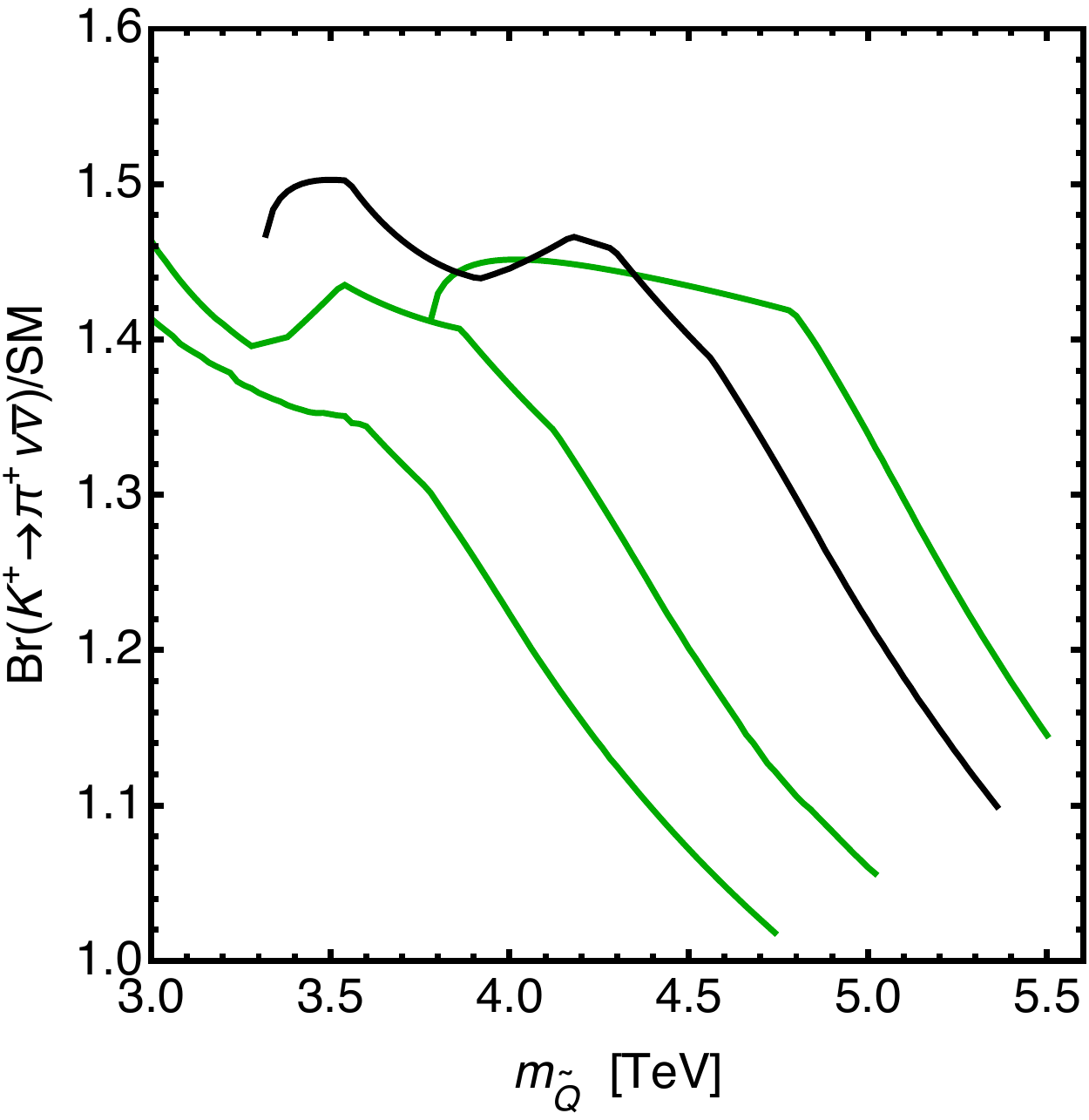}
\caption{
The maximum value of $\mathcal{B}(K^+\to\pi^+\nu\bar\nu)$ normalized by the SM prediction as a function of $m_{\tilde Q}$.
Here, $( \epoe )^{\rm SUSY} = 10.0 \times 10^{-4}$ is fixed.
The parameters are $\gamma_R/\beta_R = -\gamma_L/\beta_L = 1$ and $m_{\tilde g}/m_{\tilde Q} = 1$ on the black line. 
In the left plot, $\gamma_R/\beta_R = -\gamma_L/\beta_L = 0.6, 0.8, 1.2$ with $m_{\tilde g}/m_{\tilde Q} = 1$ from left to right of the red lines.
In the right plot, $m_{\tilde g}/m_{\tilde Q} = 1.8, 1.4, 0.8$ with $\gamma_R/\beta_R = -\gamma_L/\beta_L = 1$ from left to right of the green lines.
}
\label{fig:KPpinn}
\end{center}
\end{figure}

Next, $\mathcal{B}(K^+\to\pi^+\nu\bar\nu)$ is maximized for given $m_{\tilde Q}$ in Fig.~\ref{fig:KPpinn}.
The branching ratio depends on $\mathcal{C}_{HQ}$ and $\mathcal{C}_{HD}$ similarly to the case of $\mathcal{B}(K_L\to\pi^0\nu\bar\nu)$.
Hence, it can be larger than the SM prediction when either $\beta_L\gamma_L$ or $\beta_R\gamma_R$ is negative.
The real component of $\mathcal{C}_{HQ}$ and $\mathcal{C}_{HD}$ contributes to the ratio, which is different from the case of $\mathcal{B}(K_L\to\pi^0\nu\bar\nu)$ and \epoe. 
Consequently, the peak structure in Fig.~\ref{fig:KLpinn} disappears. 
The maximal value tends to decrease as $m_{\tilde Q}$ increases.
They are enhanced when $|\gamma_i/\beta_i|$ is small and $m_{\tilde g}/m_{\tilde Q}$ is large.
The maximal value can be about 1.6--1.7 times larger than the SM prediction.
The deviation could be measured in the current NA62 experiment.

\begin{figure}[t]
\begin{center}
\includegraphics[scale=0.5, bb= 0 0 360 356]{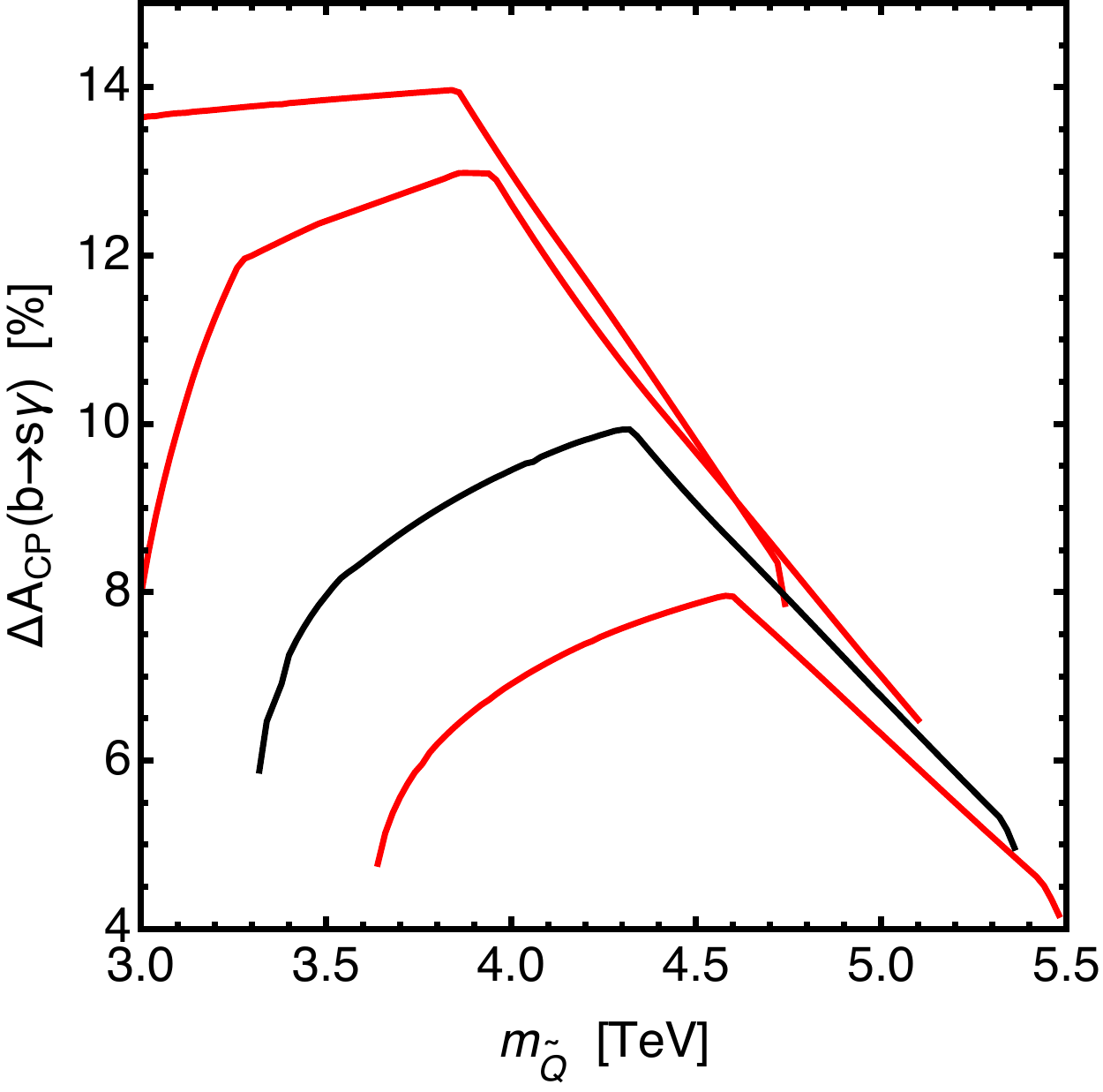}\hspace{4mm}
\includegraphics[scale=0.5, bb= 0 0 360 374]{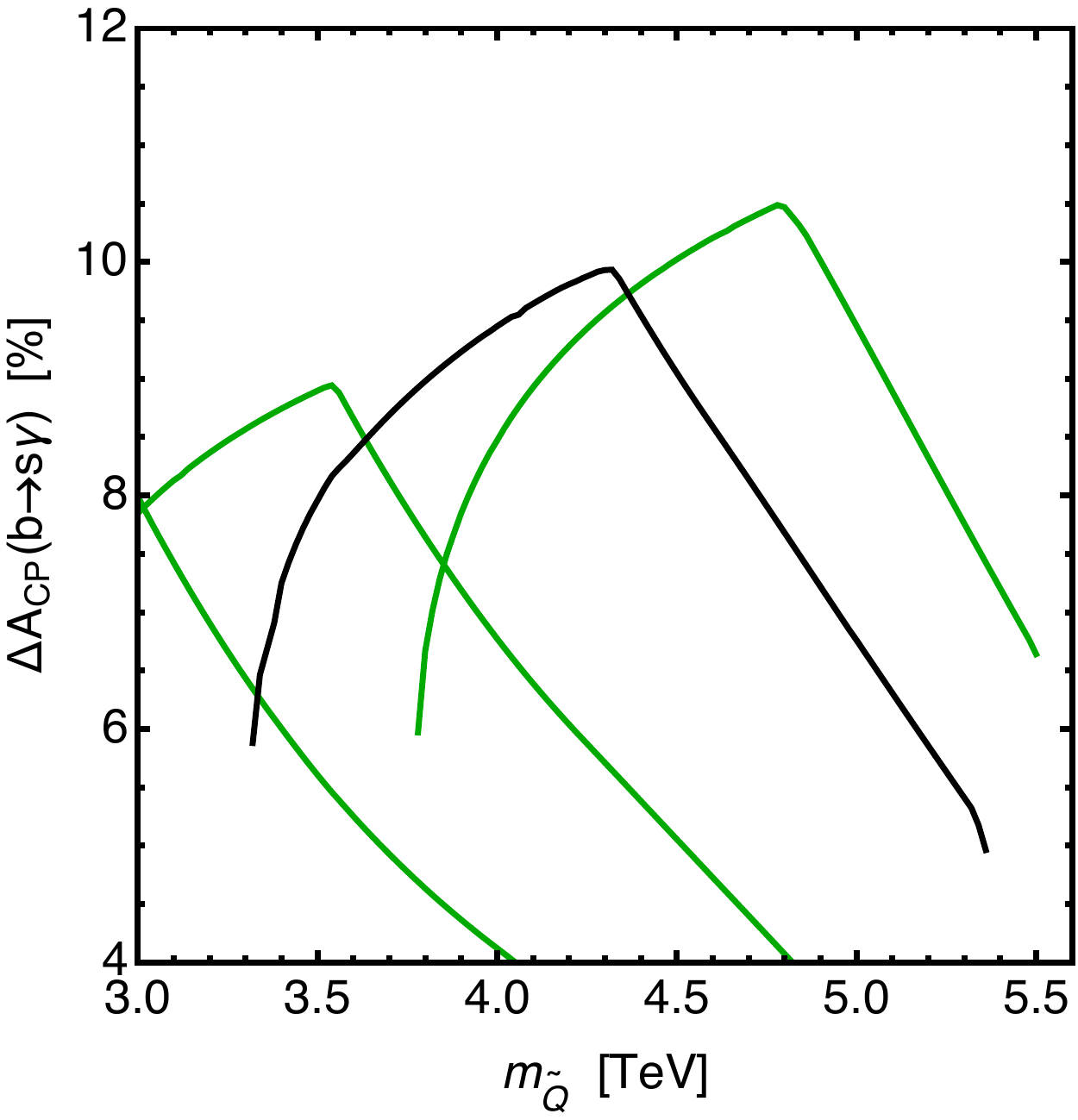}
\caption{
The maximum value of $\Delta A_{\rm CP}(b\to s\gamma)$ as a function of $m_{\tilde Q}$.
Here, $( \epoe )^{\rm SUSY} = 10.0 \times 10^{-4}$ is fixed.
The parameters are $\gamma_R/\beta_R = -\gamma_L/\beta_L = 1$ and $m_{\tilde g}/m_{\tilde Q} = 1$ on the black line. 
In the left plot, $\gamma_R/\beta_R = -\gamma_L/\beta_L = 0.6, 0.8, 1.2$ with $m_{\tilde g}/m_{\tilde Q} = 1$ from left to right of the red lines.
In the right plot, $m_{\tilde g}/m_{\tilde Q} = 1.8, 1.4, 0.8$ with $\gamma_R/\beta_R = -\gamma_L/\beta_L = 1$ from left to right of the green lines.
}
\label{fig:DAcp}
\end{center}
\end{figure}

Let us also mention about the \CPV observable, $\Delta A_{\rm CP}(b\to s\gamma)$.
In the analysis, since the \CPV phases arise in $(T_D)_{23}$ and $(T_D)_{32}$, the asymmetry can be sizable.
In Fig.~\ref{fig:DAcp}, the maximum value of $\Delta A_{\rm CP}(b\to s\gamma)$ is shown as a function of $m_{\tilde Q}$.
Here, $( \epoe )^{\rm SUSY} = 10.0 \times 10^{-4}$ is fixed.
On the black line, $\gamma_R/\beta_R = -\gamma_L/\beta_L = 1$ and $m_{\tilde g}/m_{\tilde Q} = 1$ are chosen.
In the left plot, the trilinear coupling is varied as $\gamma_R/\beta_R = -\gamma_L/\beta_L = 0.6, 0.8, 1.2$ with $m_{\tilde g}/m_{\tilde Q} = 1$ from left to right of the red lines.
In the right plot, the gluino mass is set as $m_{\tilde g}/m_{\tilde Q} = 1.8, 1.4, 0.8$ with $\gamma_R/\beta_R = -\gamma_L/\beta_L = 1$ from left to right of the green lines.
It is found that the asymmetry is enhanced especially when $|\gamma_i/\beta_i|$ is small, because smaller ratios lead to larger $(T_D)_{23}$ and $(T_D)_{32}$ to achieve $( \epoe )^{\rm SUSY} = 10.0 \times 10^{-4}$.
Also, when $m_{\tilde g}/m_{\tilde Q}$ is small, the asymmetry becomes large.
The \CP asymmetry can be as large as 14\% for $\gamma_R/\beta_R = -\gamma_L/\beta_L = 0.6$.
We also find that $\Delta A_{\rm CP}(b\to s\gamma)$ is likely to be positive when it is enhanced in our scenario. 
Such an asymmetry seems to be large enough to be measured at Belle II with $50\,{\rm ab}^{-1}$.\footnote{
  Although a part of the parameter regions seems to be constrained by the current experimental result \eqref{eq:ExpDAcp}, the theoretical uncertainty is large, and thus, we have not employed this limit.
}

\begin{figure}[t]
\begin{center}
\includegraphics[scale=0.5, bb= 0 0 360 360]{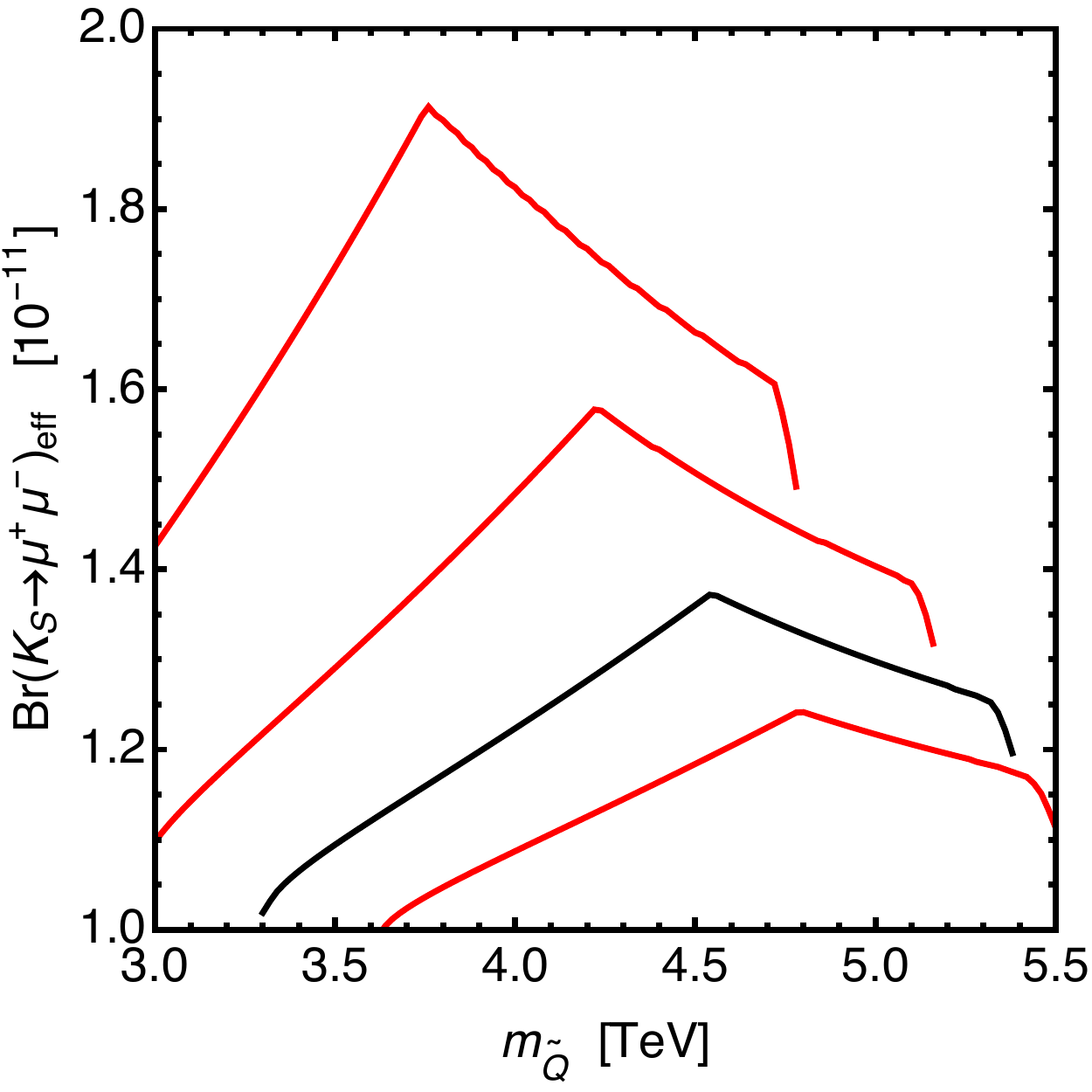}\hspace{4mm}
\includegraphics[scale=0.5, bb= 0 0 360 369]{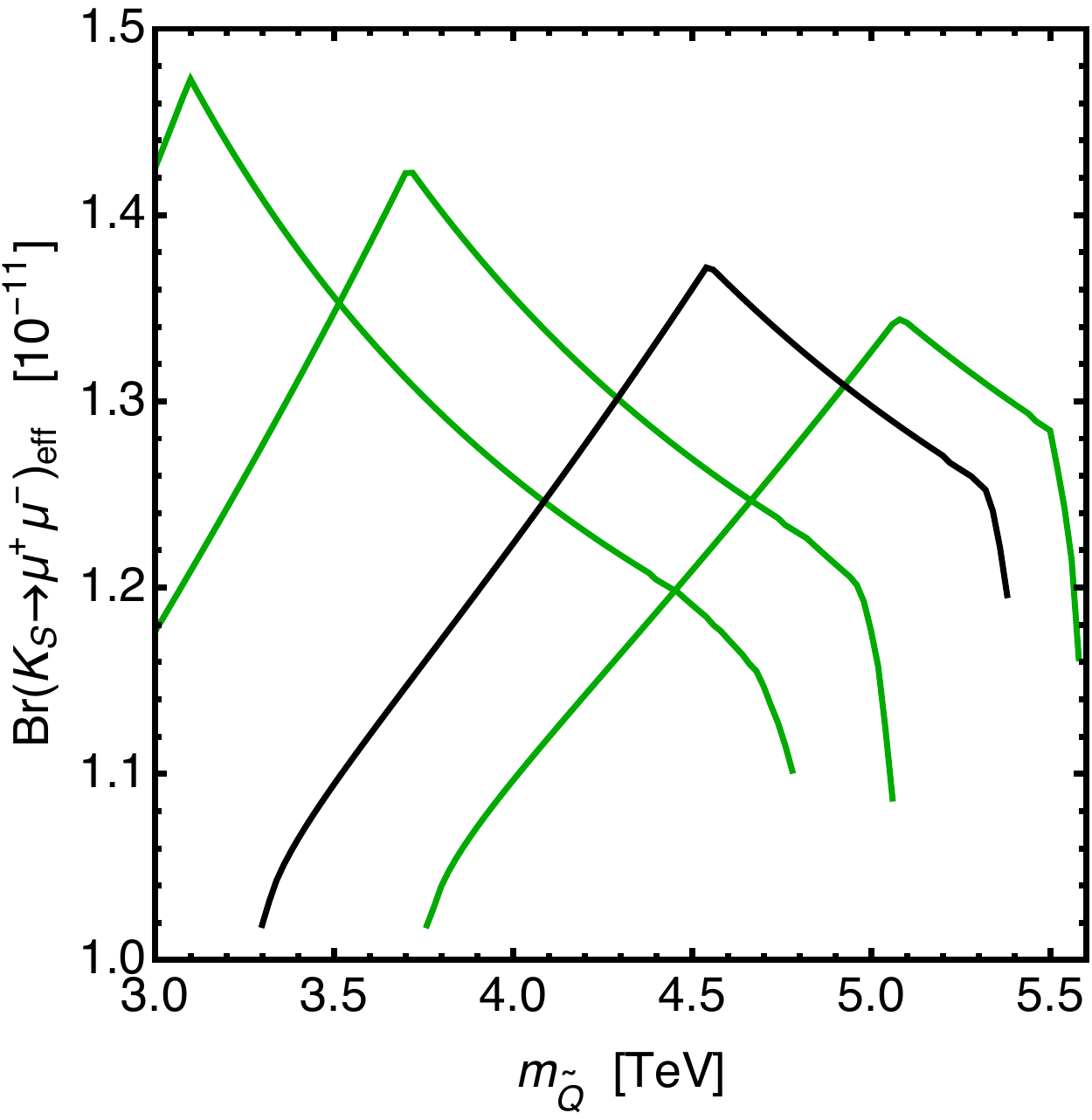}
\caption{The effective branching ratio of $ K_S \to \mu^+ \mu^-$ is shown.
Here, $D=1$ and $\eta_\mathcal{A}=-1$ are chosen.
The model parameters are the same as those in Fig.~\ref{fig:KLpinn}.
Here, $( \epoe )^{\rm SUSY} = 10.0 \times 10^{-4}$.
The parameters are $\gamma_R/\beta_R = -\gamma_L/\beta_L = 1$ and $m_{\tilde g}/m_{\tilde Q} = 1$ on the black line. 
In the left plot, $\gamma_R/\beta_R = -\gamma_L/\beta_L = 0.6, 0.8, 1.2$ with $m_{\tilde g}/m_{\tilde Q} = 1$ from left to right of the red lines.
In the right plot, $m_{\tilde g}/m_{\tilde Q} = 1.8, 1.4, 0.8$ with $\gamma_R/\beta_R = -\gamma_L/\beta_L = 1$ from left to right of the green lines.
}
\label{fig:KSmm}
\end{center}
\end{figure}

Finally, we study the SUSY contribution to  $K_S \to \mu^+\mu^-$ as a function of $m_{\tilde Q}$.
They are enhanced when the sign of the left-handed contribution is opposite to that of the right-handed one.
Such a setup is realized in this subsection.
In Fig.~\ref{fig:KSmm}, the effective branching ratio of $K_S \to \mu^+ \mu^-$ is shown.
Here, the dilution factor $D=1$ and the relative sign $\eta_\mathcal{A}=-1$ are chosen as a reference case.\footnote{
In the case of $D=0$, we find that the branching ratio $\mathcal{B}(K_S \to \mu^+ \mu^-)$ in Eq.~\eqref{eq:KSmmBr} is not deviated from the SM value \eqref{eq:KSMUMU_SMD0} sizably.
}
Since the interference term is almost independent of a real component of $\mathcal{C}_{H-}$ in the parameter regions of our interest, $\mathcal{B}\left( K_S \to \mu^+ \mu^- \right)_{\rm eff}$ is determined once $( \epoe )^{\rm SUSY}$ and $\mathcal{B}(K_L\to\pi^0\nu\bar\nu)$ are given.
Therefore, in Fig.~\ref{fig:KSmm}, we take the same $\alpha_i$, $ \beta_i$ and $\gamma_i$ as those in Fig.~\ref{fig:KLpinn}, which maximize $\mathcal{B}(K_L\to\pi^0\nu\bar\nu)$.
It is found that $\mathcal{B}\left( K_S \to \mu^+ \mu^- \right)_{\rm eff}$ is enhanced especially when $|\gamma_i/\beta_i|$ is small.
The effective branching ratio can be $1.9 \times 10^{-11}$,
which is larger than the SM prediction \eqref{eq:KSMUMU_SMD1}.
Such a branching ratio might be measured by the end of the LHCb Run-2, and it is large enough to be detected at the LHCb Run-3 \cite{LHCbupgrade}.

\section{Conclusions}
\label{sec:conclusion}

In this paper, we studied \CP violations in the neutral kaon decay in the
MSSM scenario where non-minimal flavor mixings and \CPV phases
reside in the trilinear scalar couplings of the down-type squarks.
We calculated SUSY contributions that are induced by one-loop diagrams
involving gluino and squarks, and evaluated their effects on flavor
observables.
We took the top-Yukawa contributions to $\Delta S = 2$ observables into account.
Considering constraints from the
vacuum stability and the measurements of \epsk,
$\mathcal{B}(K_L\to \mu^+\mu^-)$,
$\mathcal{B}(\bar{B}\to X_s\gamma)$ and $\mathcal{B}(\bar{B}\to X_d\gamma)$, 
we searched for the allowed parameter regions of the trilinear coupling
parameters and investigated possible effects on \epoe,
$\mathcal{B}(K_L\to \pi^0\,\nu\,\bar{\nu})$,
$\mathcal{B}(K^+\to \pi^+\,\nu\,\bar{\nu})$,
$\mathcal{B}(K_S\to\mu^+\,\mu^-)_{\rm eff}$ and
$\Delta A_{\mathrm{CP}}(b\to s\,\gamma)$.

We found that the difference between the measured value and the
SM prediction of \epoe\,can be explained by
the gluino-mediated $Z$-penguin contribution to the $s\to d$ transition
amplitude for the squark mass smaller than $5.6\TeV$.
In addition, 
$\mathcal{B}(K_L\to \pi^0\,\nu\,\bar{\nu})$ and
$\mathcal{B}(K^+\to \pi^+\,\nu\,\bar{\nu})$ can be enhanced by about 
$50\,\%$ and $70\,\%$ of the SM values, respectively.
It is also shown that $\mathcal{B}(K_S\to\mu^+\,\mu^-)_{\rm eff}$ and
$\Delta A_{\mathrm{CP}}(b\to s\,\gamma)$ are significantly enhanced.

The deviations from the SM predictions of these observables can
be probed in near-future experiments such as KOTO, NA62, LHCb and Belle II.
Since the pattern of the deviations is closely related to the structure
of the trilinear coupling matrix in the model, the measurements would 
provide us with important clues to explore
flavor structures in physics beyond the SM.

\vspace{1em}
\noindent {\it Acknowledgements}:
We are grateful to J.~A. Evans and D. Shih for helping us to compare our numerical results for some of FCNC observables to outputs from the FormFlavor code~\cite{Evans:2016lzo}. 
We would also like to thank A.~Ishikawa for valuable comments about $\Delta A_{\rm CP}(b \to s \gamma) $ in the Belle experiment.
This work was supported by JSPS KAKENHI No.~16K17681 (M.E.), 16H03991 (M.E.), 16H06492 (K.Y.) and 17K05429 (S.M.).


\end{document}